\documentclass[twocolumn]{autart}    

\usepackage{graphicx}          

\usepackage{amsfonts,amssymb,bm}
\usepackage{amsmath}
\usepackage{color}

\newtheorem{remark}{Remark}
\newtheorem{lemma}{Lemma}
\newtheorem{theorem}{Theorem}
\newtheorem{proof}{Proof}

\newtheorem{corollary}{Corollary}
\newtheorem{assumption}{Assumption}

\usepackage{booktabs}
\usepackage{algorithm}
\usepackage{algpseudocode}

\begin{document}

\begin{frontmatter}

\title{Recursive Sparse Parameter Identification of Multivariate \textcolor{black}{ARMAX Systems} with Non-stationary Observations and {\color{black}{Colored Noise}} (Extended Version) \thanksref{footnoteinfo}} 

\thanks[footnoteinfo]{This research was supported by National Key R\&D Program of China under Grant 2024YFC3307201, National Natural Science Foundation of China under Grant 12288201, and CAS Project for Young Scientists in Basic Research under Grant YSBR-008.}

\author[FU]{Yanxin Fu}\ead{fuyanxin@bit.edu.cn},    
\author[ZHAO]{Wenxiao Zhao}\ead{wxzhao@amss.ac.cn}
\address[FU]{State Key Lab of Autonomous Intelligent Unmanned Systems, Beijing Institute of Technology, Beijing 100081, China and School of Automation, Beijing Institute of Technology, Beijing 100081, China}  
\address[ZHAO]{State Key Laboratory of Mathematical Sciences (SKLMS), Academy of Mathematics and Systems Science, Chinese Academy of Sciences, Beijing 100190, China and School of Mathematical Sciences, University of Chinese Academy of Sciences, Beijing 100049, China}             


\begin{abstract}                          
The classical sparse parameter identification methods are usually based on the iterative basis selection such as greedy algorithms, or the numerical optimization of regularized cost functions such as LASSO and Bayesian posterior probability distribution, etc., which, however, are not suitable for online sparsity inference when data arrive sequentially. This paper presents recursive algorithms for sparse parameter identification of multivariate ARMA systems with exogenous inputs (ARMAX) in the presence of non-stationary observations and colored noise. First, {\em a posteriori} estimates for the system noise are given. Based on the noise estimates, the system inputs, and the system outputs, a new bivariate criterion function is presented by introducing an auxiliary variable matrix into a weighted $L_1$ regularization criterion. The new criterion function is subsequently decomposed into two solvable subproblems via alternating optimization of the two variable matrices, for which the optimizers can be explicitly formulated into recursive equations. Second, under the non-stationary and non-persistent excitation conditions in the input-output observations, the consistency of parameter estimation and sparsity recovery of multivariate ARMAX systems with the embedded colored noise model is established. That is, the estimates are proved to be with (i) set convergence, i.e., the accurate estimation of the sparse index set of the unknown parameter matrix, and (ii) parameter convergence, i.e., the consistent estimation for values of the non-zero elements of the unknown parameter matrix. 
Finally, numerical examples are given to support the theoretical analysis.
\end{abstract}

\date{}

\end{frontmatter}

\section{Introduction}
Sparsity of complex models has many manifestations in reality, including but not limited to audio signal representations in frequency domains, transient impulse responses in time domains \cite{nagahara2024survey}, topology structure of networked systems \cite{ZORZI2022Nonparametric}, etc. 
Due to the inherent advantages of sparse models in achieving data compression, computational efficiency, and model interpretability, the dual challenges of reconstructing sparse structures and accurately estimating the values of non-zero entries in parameter matrices of stochastic systems have become an important research frontier. Recent advances span multiple domains, for example, compressed sensing in signal processing \cite{ben2010coherence,candes2005decoding,candes2008enhancing,chartrand2008iteratively,chen2001atomic,donoho2012sparse,needell2009cosamp,obermeier2017sensing,wang2020accelerated}, variable selection in statistics \cite{fan2001variable,friedman2012fast,park2008bayesian,tibshirani1996regression,zhang2010analysis,zou2006adaptive}, feature selection and neural network pruning in deep learning \cite{axiotis2021sparse,lemhadri2021lassonet,liu2020dynamic,scardapane2017group,tang2022survey}, and sparse identification in systems and control \cite{brunton2016discovering,fu2023support,guo2024sparse,parsa2023transformation,pillonetto2022sparse,zhao2020sparse,zhou2022sparse}.

As far as the methodological paradigms for sparse parameter estimation are of concern, three classes of methods, i.e., greedy algorithms, regularization techniques, and Bayesian inference framework, have been systematically developed from a computational modeling viewpoint. The greedy methods, exemplified by orthogonal matching pursuit (OMP) \cite{ben2010coherence,donoho2012sparse,parsa2023transformation}, iterative hard thresholding \cite{axiotis2021sparse}, and compressive sampling matching pursuit \cite{needell2009cosamp}, iteratively select the most relevant basis elements, progressively update residuals, and efficiently construct sparse solutions through a sequential approximation process. While the greedy methods are easy to calculate through heuristic basis selection, their accuracy decreases significantly in high-dimensional problems \cite{wang2020accelerated}. The regularization-based approaches involve an optimization problem with an $L_0$ non-convex penalty, which is often relaxed to an $L_1$ convex surrogate. The so-called basis pursuit denoising (BPDN) method \cite{chen2001atomic}, least absolute shrinkage selection operator (LASSO) method \cite{tibshirani1996regression}, and sparse identification of nonlinear dynamics (SINDy) method \cite{brunton2016discovering} belong to this category. These approaches are further refined through reweighted $L_1$ regularization schemes \cite{candes2008enhancing,chartrand2008iteratively}, conceptually similar to the adaptive LASSO method \cite{zou2006adaptive}, wherein adaptive weights are incorporated into the $L_1$ regularization term to enhance sparsity recovery performance. Within the category of regularization-based approaches, the non-convex regularization strategies are also developed, including $L_p$ norm penalty with $0<p<1$ \cite{guo2024sparse}, smoothly clipped absolute deviation (SCAD) penalty \cite{fan2001variable}, logarithm penalty \cite{friedman2012fast}, minimax concave penalty \cite{zhang2010nearly}, and capped $L_1$ penalty \cite{zhang2010analysis}, etc. As for the Bayesian methods, they basically treat the unknown parameters as random variables governed by probabilistic distributions \cite{park2008bayesian,pillonetto2022sparse,zhou2022sparse}. This framework incorporates sparsity-inducing priors with learnable hyper-parameters, notably the Laplace prior distribution \cite{park2008bayesian}, the stable spline horseshoe prior distribution \cite{pillonetto2022sparse}, the group sparsity-inducing prior structure information \cite{zhou2022sparse}, etc. Despite demonstrating the superiority in sparsity recovery accuracy, Bayesian methods present some limitations due to the computational complexity in high-dimensional posterior distributions and the inherent sensitivity to prior specifications.

For theoretical analysis of the above methods, technical conditions such as the restricted isometry property (RIP) \cite{candes2005decoding}, or mutual coherence criteria \cite{obermeier2017sensing}, or stationarity on the observations \cite{zou2006adaptive} are often required. However, such conditions become inapplicable in identification of control systems, where the current inputs depend on past inputs, outputs, system noises, and possibly exogenous excitations; accordingly, the observation sequences are non-stationary in general. As far as the algorithm design is of concern, the greedy methods, the regularization techniques, and the Bayesian inference framework are iterative or numerical optimization-based algorithms in nature. When new observations are available, the algorithms need to be reperformed with the entire data set to obtain the new estimates. This is time-consuming for the online sparsity inference, particularly for dynamic control systems. 

For the sparse parameter identification with non-stationary observations, theoretical and algorithmic advances have been made recently. For example, \cite{zhao2020sparse} introduces a weighted $L_1$ regularization approach, and under the non-stationary and non-persistent excitation condition the paper proves that the estimates achieve both \emph{set convergence} (exact sparsity recovery of the unknown parameter matrix) and \emph{parameter convergence} (consistent estimates for nonzero entries of the unknown parameter matrix). Several extensions from \cite{zhao2020sparse} are reported in literature, c.f., the weighted $L_p~(0<p<1)$ regularization method for sparse parameter identification in \cite{guo2024sparse}, the distributed algorithms in \cite{gan2023distributed}, the algorithms and theory for sparsity inference of colored noise model with non-Gaussian distribution in \cite{fu2023support}, and the row-grouped structural sparsity recovery in \cite{zhang2023multi}. Note that in control systems highly correlated regression matrices are often involved, which undermine performance of OMP. In a recent paper \cite{parsa2023transformation}, a transformation framework for regressors that enforces low mutual coherence in the transformed domain is presented, which in turn enables exact sparsity recovery via OMP under known sparsity levels. However, the above algorithms for sparsity inference of control systems are essentially off-line, which requires high computational cost when data arrives sequentially. 
On the other hand, online/recursive algorithms for sparsity recovery has a wide range of practical scenarios, such as the space-time adaptive processing (STAP) in radar research \cite{Sun2016}, beam adaptive tracking in millimeter-wave MIMO systems within the field of wireless communication \cite{Gao2016}\cite{Mehrotra2023}, structural damage assessment \cite{Huang2017}, etc. This generates a practical demand for the design of recursive sparsity recovery algorithms and the analysis of their theoretical properties. To the authors' knowledge, the corresponding theoretical research, particularly, with non-stationary observations, has not yet been addressed in literature.

In this paper, we investigate the recursive algorithms for sparse parameter identification of multivariate ARMAX systems in the presence of non-stationary observations and colored noise. 
The contributions of the paper are given below. 
\begin{itemize}
    \item \textcolor{black}{First, we propose recursive sparsity recovery algorithms tailored to multivariate ARMAX systems with embedded colored system noise.} With the {\em a posteriori} estimates of the system noise, we first introduce a convex optimization-based criterion with an adaptive $L_1$ regularization term for sparsity recovery. Leveraging the alternating optimization framework for distinct variable sets, we introduce an auxiliary variable matrix into the weighted \(L_1\) regularization criterion, thereby deriving a bivariate optimization criterion in a higher-dimensional space. The new criterion is subsequently decomposed into two solvable subproblems via alternating optimization of the two variable matrices, for which the optimizers can be explicitly formulated into recursive equations. 
    \item Second, we establish the consistency of parameter estimation and sparsity recovery of multivariate ARMAX systems with the embedded colored noise model, under the conditions of non-stationarity and non-persistent excitation in the input-output observations. That is, the estimates are proved to be with (i) set convergence, i.e., the sparse index set of the unknown parameter matrix is correctly identified with a finite number of observations, and (ii) parameter convergence, i.e., the estimates for values of the non-zero elements of the unknown parameter matrix converge to the true values with probability one as the number of observations tends to infinity. To the authors' knowledge, the excitation condition introduced in this paper is possibly the weakest condition among existing literature for sparse parameter identification with non-stationary observations. 
\end{itemize}



The rest of the paper is organized as follows. Section \ref{sec:onalgarx} presents the recursive sparse identification algorithms for multivariate ARMAX systems. \textcolor{black}{Sections \ref{sec:alg1theory} and \ref{sec:alg1theory0} present theoretical results} 
followed by numerical validation in Section \ref{sec:sim}.

\textbf{Notations:} Let $(\Omega, \mathcal{F}, \mathbb{P})$ be the probability space and $\omega$ be an element in $\Omega$. Denote by $\mathbb{E}[\cdot]$ the mathematical expectation operator. Let $\|\cdot\|_1$ denote the $1$-norm of vectors or matrices. Denote the Frobenius norm of vectors and matrices by $\|\cdot\|$. For two positive sequences $\{a_k\}_{k \geq 1}$ and $\{b_k\}_{k \geq 1}$, by $a_k = O(b_k)$ it means $a_k \leq cb_k,~k \geq 1$ for some $c > 0$ and by $a_k = o(b_k)$ it means $a_k/b_k \rightarrow 0$ as $k \rightarrow \infty$. Let ${\rm sgn}(x)$ be the sign function, i.e., ${\rm sgn}(x)=1$ if $x \geq 0$ and ${\rm sgn}(x)=-1$ if $x < 0$. For a matrix $D$, its $\left(s,t\right)$-th entry is denoted by $D(s,t)$ and its $s$-th column is denoted by $D(s)$. For a vector $d\in \mathbb{R}^m$, without confusion in notations, its $l$-th entry is also denoted by $d(l)$. For a symmetric matrix $D$, its maximal and minimal eigenvalues are denoted by $\lambda_{\max}\{D\}$ and $\lambda_{\min}\{D\}$, respectively. Denote by $\otimes$ the Kronecker product of matrices. For a set $\mathcal{A}$, denote its complement set as $\mathcal{A}^c$.

\section{Recursive Algorithms for Sparse Parameter Identification}\label{sec:onalgarx}

Consider the parameter identification of the following multivariate ARMAX system, 
\begin{align}
A(z)y_{k+1}=B(z)u_k+C(z)w_{k+1},~ k \geq 0,\label{eq:ARMAX}
\end{align} 
where the matrix polynomial $A(z)=I_n+A_1z+\cdots+A_pz^p$, $B(z)=B_1+B_2z+\cdots+B_qz^{q-1}$, $C(z)= I_n+C_1z+\cdots+C_rz^r$, 
$z$ is the back-shift operator, i.e., $z y_{k+1}=y_k$, $A_i,~i=1,\cdots,p$, $B_j,~j=1,\cdots,q$, and $C_\nu,~\nu=1,\cdots,r$ are unknown parameter matrices, $p>0,~q>0,~r>0$ are known upper bounds of system orders, $u_k\in \mathbb{R}^l,~ y_k\in \mathbb{R}^n,~ w_k\in \mathbb{R}^n$ are the system inputs, outputs, and the system noise, respectively. 

Denote $d \triangleq np+lq+nr$,   
$$
\Theta \triangleq [-A_1,\cdots, -A_p, B_1, \cdots, B_q, C_1, \cdots, C_r]^\top\in \mathbb{R}^{d \times n},
$$
and 
\begin{align}
\varphi_k^0=[&y_k^\top, \cdots, y_{k-p+1}^\top, u_k^\top, \cdots, u_{k-q+1}^\top,\nonumber\\ 
&w_k^\top ,\cdots, w_{k-r+1}^\top]^\top\in \mathbb{R}^{d}.\nonumber
\end{align} 
Then the ARMAX system (\ref{eq:ARMAX}) can then be reformulated in regression form as:
\begin{align}
y_{k+1}=\Theta^\top \varphi^0_k + w_{k+1},~~k \geq 0,\label{eq:yk}
\end{align}
where $\Theta\in \mathbb{R}^{d\times n}$ denotes the unknown parameter matrix to be identified, and $\varphi^0_k \in \mathbb{R}^{d}$ is the regression vector that incorporates both the system inputs and outputs, as well as the unknown system noise.

Define a sequence of $\sigma$-algebras $\{\mathcal{F}_k\}_{k\geq 1}$ as 
$
\mathcal{F}_k \triangleq \sigma \left\{y_{k},\cdots,y_0,u_{k-1},\cdots,u_0,w_k,\cdots,w_1\right\}. 
$ 
In this paper, by non-stationarity it means that $\varphi^0_k\in \mathcal{F}_k,~k\geq 1$ and thus the system admits feedback control. Using non-stationary observations $\{u_k,y_{k+1}\}_{k\geq1}$, the task of sparse parameter identification is to correctly identify the set of the zero entries in $\Theta$ and to estimate the values of the nonzero entries in $\Theta$. 

For a given matrix $X \in \mathbb{R}^{d\times n}$, denote its sparse index set by
\begin{align}\label{eq:AN}
\!\!\!\mathcal{A}(X) \!\triangleq\! \{(s,t)|X(s,t)\!=\!0,s=1,\!\cdots\!,d,t=1,\!\cdots\!,n\}.
\end{align}
Define
\begin{align}\label{eq:A}
\mathcal{A}^*\triangleq \mathcal{A}(\Theta)
\end{align}
and its complement set by ${\mathcal{A}^*}^c$. 

Denote
$
Y_{N+1}=
\left[
y_2,
\cdots,
y_{N+1}
\right]^{\top}\in \mathbb{R}^{N \times n},~
\Phi_N^0=
\left[
\varphi_1^0,
\cdots,
\varphi_{N}^0
\right]^{\top}\in \mathbb{R}^{N \times d},~
\mathcal{W}_{N+1}=
\left[
w_2,
\cdots,
w_{N+1}
\right]^{\top}\in \mathbb{R}^{N \times n}$.
For system (\ref{eq:yk}), it follows that
$
Y_{N+1}=\Phi_N^0 \Theta+\mathcal{W}_{N+1}.
$ 
Note that $\varphi^0_k$ incorporates the system noise and cannot be directly applied to the algorithm design. We first introduce {\em a posteriori} estimates for the noise (called the over-parameterization technique in literature, see, e.g., \cite{chen2012identification}). Based on the noise estimate, we then construct a regression vector $\varphi_k$ available for the algorithm design. 
After that, we develop an algorithm that promotes sparsity in the estimation. 

We assume that the vector \(\varphi_k\) (required for algorithm design) is available, with the detailed algorithm provided below. In the regularization techniques for sparse parameter identification of system (\ref{eq:yk}), the estimates are usually generated by optimization of the following cost function \cite{zhao2020sparse,zou2006adaptive},
\begin{align}\label{eq:bivariate00}
\!\!\!\min_{X \in \mathbb{R}^{d\times n}}\!\! \frac{1}{2}\|Y_{N+1}-\Phi_N X\|^2\!+\!\sum_{s=1}^d \sum_{t=1}^n \! \gamma_N(s,t) \left|X(s,t)\right|
\end{align}
where 
$
\Phi_N=
\left[
\varphi_1,
\cdots,
\varphi_{N}
\right]^{\top}\in \mathbb{R}^{N \times d}
$ 
and $\gamma_N(s,t),~s=1,\cdots,d,~t=1,\cdots,n$ are the weights for the $L_1$ penalty term.

Note that the above optimization problem (\ref{eq:bivariate00}) is convex but non-smooth, whose solution has no explicit form in general and relies on numerical iterative methods. When the number of observations increases, c.f., from $N$ to $N+1$, to obtain a new estimate one needs to re-optimize the cost function with the whole data set $\{\varphi_k,y_{k+1}\}_{k=1}^{N+1}$. This is time-consuming for online parameter identification. In the following, an auxiliary variable matrix is introduced into the regularization criterion (\ref{eq:bivariate00}) to obtain a bivariate optimization criterion in a higher dimensional space. The new criterion is subsequently decomposed into two more easily solvable subproblems via alternating optimization of the two variable matrices, for which the optimizers can be explicitly formulated into recursive equations.

With an auxiliary variable matrix $\Xi\in \mathbb{R}^{d\times n}$, a bivariate optimization problem in $\mathbb{R}^{2(d\times n)}$ is given by
\begin{align}\label{eq:bivariate0}
\min_{X,\Xi\in \mathbb{R}^{d\times n}} \frac{1}{2}\|Y_{N+1}-\Phi_N X\|^2&+\sum_{s=1}^d \sum_{t=1}^n \gamma_N(s,t) \left|\Xi(s,t)\right|\nonumber\\
&+\frac{\mu}{2}\|X-\Xi\|^2
\end{align}
where $\mu>0$ is a tuning parameter and the quadratic term $\frac{\mu}{2}\|X-\Xi\|^2$ balances the estimation error while preserving the sparsity of the estimates.

With arbitrary initial values $\Theta_{0}$ and $\Xi_{0}$, for online sparse parameter identification, i.e., the situation that the number of observations $N$ increases, the following alternating minimization approach for solving the optimization problem (\ref{eq:bivariate0}) is applied,
\begin{align}
\Theta_{N+1} &\!= \!\mathop{{\rm argmin}} \limits_{X\in \mathbb{R}^{d \times n}} \frac{1}{2}\|Y_{N+1}\!-\!\Phi_N X\|^2\!+\!\frac{\mu}{2}\|X-\Xi_N\|^2,\label{pro0:theta}\\
\Xi_{N+1} &= \mathop{{\rm argmin}} \limits_{\Xi \in \mathbb{R}^{d \times n}} \frac{\mu}{2}\|\Theta_{N+1}-\Xi\|^2\nonumber\\
&~~~~~~~~~~~~~~~+ \sum_{s=1}^d \sum_{t=1}^n \gamma_N(s,t) \left|\Xi(s,t)\right|,\label{pro0:xi}
\end{align}
for all $N\geq0$.

\begin{remark}\label{Rmk:1}
The design philosophy underlying Algorithms (\ref{pro0:theta})--(\ref{pro0:xi}) differs from that of alternating direction method of multipliers (ADMM) \cite{boyd2011distributed,wang2013online}; however, they share the fundamental principle of alternating optimization. For a general ADMM procedure, it is formulated as the augmented Lagrangian with a dual variable of the original optimization problem. 
In this paper, without adopting the dual variable, we introduce an auxiliary variable matrix into the regularization criterion (\ref{eq:bivariate00}) to derive a bivariate optimization criterion (\ref{eq:bivariate0}) with respect to variables $X$ and $\Xi$. The benefits are twofold: First, the dimension of the optimization space is lower than that of ADMM, since the dual variable in ADMM is not adopted; Second, based on the specific design of (\ref{eq:bivariate0}), algorithms (\ref{pro0:theta})--(\ref{pro0:xi}) have closed-form solutions which can be formulated into recursive equations when new data arrives online. \textcolor{blue}{In essence, the proposed framework, given by (\ref{pro0:theta})--(\ref{pro0:xi}), belongs to the category of proximal alternating minimization (PAM) methods \cite{attouch2010proximal}. Compared with the classical PAM methods, which are generally offline algorithms, the framework given by (\ref{pro0:theta})--(\ref{pro0:xi}) enables an online computational variant.}
\end{remark}

Define $P_{N+1}\triangleq (\mu I_d+\Phi_N^\top \Phi_N)^{-1},~N\geq 0$ and $P_{0}\triangleq (\mu I_d)^{-1}$, where $\mu>0$ is the tuning parameter given in (\ref{eq:bivariate0}). For the algorithm (\ref{pro0:theta}), the following recursive equations hold for $\{\Theta_{N}\}_{N\geq1}$.

\begin{lemma}\label{lem:rls}
The estimates $\{\Theta_{N}\}_{N\geq1}$ generated by (\ref{pro0:theta}) have the following recursive form,
\begin{align}
\Theta_{N+1}&=\Theta_N+a_N P_N\varphi_N\left(y_{N+1}^\top-\varphi_N^\top \Theta_N\right)\nonumber\\
&~~~~+\mu P_{N+1} (\Xi_N-\Xi_{N-1}),\label{eq:thetanrls3}\\
a_N &= \left(1 + \varphi_N^\top P_N \varphi_N\right)^{-1},\label{eq:thetanrls1}\\
P_{N+1}&=P_N-a_N P_N \varphi_N \varphi_N^\top P_N.\label{eq:thetanrls2}
\end{align}
\end{lemma}

\begin{proof}
The detailed proof is omitted since it can be similarly obtained as the recursive formula of least squares algorithm. 
\end{proof}

Note that the cost functions (\ref{eq:bivariate0}) and (\ref{pro0:xi}) are weighted $L_1$ regularizations. With $\{\Theta_{N}\}_{N\geq1}$ generated from (\ref{eq:thetanrls3})--(\ref{eq:thetanrls2}), we set
\begin{align}\label{eq:thetahat}
\widehat{\Theta}_{N+1}(s,t)=&\Theta_{N+1}(s,t)+{\rm sgn}\left(\Theta_{N+1}(s,t)\right)\nonumber\\
&\cdot \sqrt{\frac{\log\lambda_{\max}(N)}{\lambda_{\min}(N)}},~~N\geq1,
\end{align}
where $\lambda_{\max}(N) \triangleq \lambda_{\max}\{P_{N+1}^{-1}\}=\lambda_{\max}\{\mu I_d+\sum_{k=1}^{N} \varphi_k \varphi_k^\top\}$ and $\lambda_{\min}(N) \triangleq \lambda_{\min}\{P_{N+1}^{-1}\}=\lambda_{\min}\{\mu I_d+\sum_{k=1}^{N} \varphi_k \varphi_k^\top\}$. For any $s=1,\cdots,d,~t=1,\cdots,n$, define
\begin{align}
\gamma_N(s,t)=\frac{\lambda_N}{\big|\widehat{\Theta}_{N+1}(s,t)\big|},\label{eq:thetahat'}
\end{align}
where $\{\lambda_N\}_{N\geq 1}$ is a positive sequence to be specified later. The sequences $\{\gamma_N(s,t),~s=1,\cdots,d,~t= 1,\cdots,n\}_{N\geq1}$ serve as the weights for each term in the $L_1$ penalty.

We now derive the recursive equation for (\ref{pro0:xi}), prefaced by the following technical lemma.

\begin{lemma}(Soft-thresholding operator, \cite{fornasier2010theoretical})\label{lem:closedsols}
For any given $\gamma>0$ and matrix $Y \in \mathbb{R}^{d \times n}$, the optimal solution to the following problem 
$
\min_{X\in \mathbb{R}^{d\times n}} \frac{1}{2}\|Y-X\|^2+\gamma \|X\|_1,
$ 
is given by the element-wise soft-thresholding operator 
$
X^*=\mathcal{S}_{\gamma}({Y}),
$ 
where 
$
\mathcal{S}_{\gamma}(Y)= \left[\mathcal{S}_{\gamma}(Y(s,t))\right],s=1,\cdots,d,t =1,\cdots,n,
$ 
with 
$
\mathcal{S}_{\gamma}(Y(s,t))={\rm sgn}(Y(s,t)) \cdot \max\{|Y(s,t)|-\gamma,0\}.
$ 
\end{lemma}

By Lemma \ref{lem:closedsols}, we obtain the closed-form solution of the sub-problem (\ref{pro0:xi}),
\begin{align}\label{eq:sparseso}
&\Xi_{N+1}(s,t)={\rm sgn}\left(\Theta_{N+1}(s,t)\right)\nonumber\\
&\cdot \max\bigg\{\left|\Theta_{N+1}(s,t)\right|-\frac{\lambda_N}{\mu|\widehat{\Theta}_{N+1}(s,t)|},0\bigg\}.
\end{align}

Combining (\ref{eq:thetanrls3})-(\ref{eq:thetanrls2}), (\ref{eq:thetahat'}), and (\ref{eq:sparseso}), we obtain a group of recursive algorithms for sparse parameter identification of system (\ref{eq:yk}). See Algorithm \ref{Alg:1} for details. 

\begin{algorithm}[hbpt!]
  \caption{Recursive Algorithms for Sparse Parameter Identification}
  \footnotesize
  \label{Alg:1}
  \begin{algorithmic}
    \Require Choosing integers $\overline{p}>p$, $\overline{q}>q$, $\overline{r}>r$, initial values $\mu>0,~P_0=\frac{1}{\mu} I_d,\overline{P}_0=\frac{1}{\mu} I_{\overline{p}n+\overline{q}l+\overline{r}n},~\alpha_0=0,~\Theta_0=0,~\Xi_0=\Xi_{-1}=0,~u_j=0,~y_j=0,~\widehat{w}_j=0,~j\leq 0$, $\{\lambda_N\}_{N\geq 1},~\lambda_N>0$
    \For{$N=0,1,2,\cdots$}
    \State \textbf{Step 1. {\em A Posteriori} Estimates of $w_N$}
    \State $\psi_{N-1}=\left[y_{N-1}^\top \cdots y_{N-\overline{p}}^\top~ u_{N-1}^\top \cdots u_{N-\overline{q}}^\top ~ {\widehat{w}}_{N-1}^\top \cdots {\widehat{w}}_{N-\overline{r}}^\top\right]^\top$
    \State $b_{N-1} = \left(1 + \psi_{N-1}^\top \overline{P}_{N-1} \psi_{N-1}\right)^{-1}$
    \State $\overline{P}_{N}=\overline{P}_{N-1}-b_{N-1} \overline{P}_{N-1} \psi_{N-1} \psi_{N-1}^\top \overline{P}_{N-1}$
    \State $\alpha_{N}=\alpha_{N-1}+b_{N-1} \overline{P}_{N-1} \psi_{N-1}\left(y_{N}^\top-\psi_{N-1}^\top \alpha_{N-1}\right)$
    \State ${\widehat{w}}_{N}=y_{N}-\alpha_{N}^\top\psi_{N-1}$
    \State \textbf{Step 2. Recursive Identification of $\Theta$}
    \State
    $\varphi_{N}=\left[y_{N}^\top \cdots y_{N-p+1}^\top~ u_{N}^\top \cdots u_{N-q+1}^\top ~ {\widehat{w}}_{N}^\top \cdots {\widehat{w}}_{N-r+1}^\top\right]^\top$
    \State
    $a_N = \left(1 + \varphi_N^\top P_N \varphi_N\right)^{-1}$
    \State
    $P_{N+1}=P_N-a_N P_N \varphi_N \varphi_N^\top P_N$
    \State $\Theta_{N+1}=\Theta_N+a_N P_N\varphi_N\left(y_{N+1}^\top-\varphi_N^\top \Theta_N\right)$
    \State $~~~~~~~~~~~~~~~~~+\mu P_{N+1} (\Xi_N-\Xi_{N-1})$
    \State \textbf{Step 3. Sparse Estimates of $\Theta$}
    \State $\widehat{\Theta}_{N+1}(s,t)=\Theta_{N+1}(s,t)$
    \State $~~~~~~~~~~~~~~~~~+{\rm sgn}\left(\Theta_{N+1}(s,t)\right) \sqrt{\frac{\log\lambda_{\max}(N)}{\lambda_{\min}(N)}}$
    \State \For{$s=1,\cdots,d,~t=1,\cdots,n$}
    \State $\Xi_{N+1}(s,t)={\rm sgn}\left(\Theta_{N+1}(s,t)\right)$
    \State $~~~~~~~~~~~~~~~~~\cdot \max\left\{\left|\Theta_{N+1}(s,t)\right|-\frac{\lambda_N}{\mu|\widehat{\Theta}_{N+1}(s,t)|},0\right\}$
    \EndFor
    \EndFor
  \end{algorithmic}
\end{algorithm}

Define an index set $\widehat{\mathcal{A}}_{N+1}\triangleq \mathcal{A}\left(\Xi_{N+1}\right)$ and a matrix sequence $\{S_{N+1}\in \mathbb{R}^{d\times n},N\geq1\}$ with
\begin{align}\label{eq21}
S_{N+1}(s,t) = \left\{
\begin{smallmatrix}
\Theta_{N+1}(s,t), & \text{ if } (s,t)\in \widehat{\mathcal{A}}_{N+1}^c,\\
0, & \text{ if } (s,t)\in \widehat{\mathcal{A}}_{N+1},
\end{smallmatrix}
\right.
\end{align}
for $s=1,\cdots,d,~t=1,\cdots,n.$ 
The zero and nonzero entries of $S_{N+1}$ serve as the estimates for the sparse entries and the nonzero entries of $\Theta$, respectively. 


\begin{remark}\label{Rmk:2}
{\color{black}{In fact, $\Xi_{N+1}$ can also be incorporated into the design of $S_{N+1}$. Set 
\begin{align*}
S_{N+1}(s,t) = \left\{
\begin{smallmatrix}
\Xi_{N+1}(s,t), & \text{ if } (s,t)\in \widehat{\mathcal{A}}_{N+1}^c,\\
0, & \text{ if } (s,t)\in \widehat{\mathcal{A}}_{N+1},
\end{smallmatrix}
\right.
\end{align*}
for $s=1,\cdots,d,~t=1,\cdots,n.$ The conclusions of Theorems 1–4 and Corollaries 1-2 presented in the following sections remain valid.}}
\end{remark}

\begin{remark}\label{Rmk:2'}
{\color{blue}{For solving the $L_1$ regularized cost function, a recursive identification algorithm is introduced in \cite{li2020online}. Compared with \cite{li2020online}, several substantial differences are made in this paper. First, {\em a more general system model} (extended from white-noise linear systems to multivariate ARMAX systems with colored noise) is considered in the paper. Second, {\em an enhanced identification algorithm tailored for sparse parameter inference} is proposed, where the {\em a posteriori} estimates for the colored noise model are first adopted and then the adaptive weights are applied to each $L_1$ term. Note that adaptive weights applied to each $L_1$ term is not considered in \cite{li2020online}. Third, this paper establishes {\em strengthened theoretical results}. In \cite{li2020online} it proves that the estimates converge to the values of the unknown parameter matrix $\Theta$ as the number of observations tends to infinity. In this paper, in addition to the consistency for estimating $\Theta$, the accuracy in sparse inference is also established, that is, the sets of the zero and the nonzero entries of $\Theta$ can be correctly identified with a finite number of observations. Fourth, compared with \cite{li2020online}, {\em the excitation conditions imposed on the input-output data sequence} have been significantly relaxed. The excitation conditions required are significantly weaker than those in \cite{li2020online} and other existing literature. See Table \ref{Tab:0} in Section \ref{sec:alg1theory}.  }}
\end{remark}

\section{Theoretical Properties of Recursive Algorithms with Conditions on ${\bf\{\varphi_k\}_{k=1}^N}$}\label{sec:alg1theory}

We introduce the following assumptions and lemma for theoretical analysis.

\begin{assumption}\label{assum:1}
The system noise $\!\{w_k,\mathcal{F}_k\}_{k \geq 1}\!$ is a martingale difference sequence (m.d.s.), i.e., $\mathbb{E}[w_{k+1}| \mathcal{F}_k]=0,$ $k \geq 1,$ and there exists some $\beta > 2$ such that ${\rm \sup}_k \mathbb{E}\big[\big.\left\|w_{k+1}\right\|^{\beta}\big| \mathcal{F}_k\big]< \infty,~ \mathrm{a.s.}$
\end{assumption}

\begin{assumption}\label{assum:2}
For each $k \geq 1$, $\varphi_k$ is $\mathcal{F}_k$-measurable.
\end{assumption}

\begin{assumption}\label{assum:cz}
    ${\rm det} C(z) \neq 0, \forall z: |z|\leq 1.$
\end{assumption}

\begin{remark}\label{rm:cz} 
{\color{black}{Note that for identification of the colored noise $C(z)w_k$, a widely applied condition in literature is the {\em strictly positive realness} (SPR) condition (see, e.g., \cite{chen2012identification}), i.e., $C^{-1}(e^{i \lambda})+C^{-\top}(e^{-i \lambda})>I, \forall \lambda \in [0,2 \pi].$ However, the SPR condition is restrictive since integrating both sides of the above inequality from $0$ to $2\pi$, we obtain $\|[C_1, \cdots, C_r]\|<1,$ which in the scalar case means $\sum_{i=1}^r C_i^2 <1.$ Assumption \ref{assum:cz} is significantly weaker than the SPR condition. Moreover, Assumption \ref{assum:cz} yields that $C^{-1}(z)$ has the following expansion
$C^{-1}(z) = \sum_{i=0}^{\infty} D_i z^i, \forall z:|z| \leq 1$ with $\|D_i\| \leq K \lambda^i,~i \geq 0$ for some $K>0$ and $\lambda \in (0,1)$.}}
\end{remark}

For the {\em a posteriori} estimates of $\{w_k\}_{k\geq0}$, it holds that
\begin{lemma}\label{lem:noise}(Theorem 4. 7 in \cite{chen2012identification})
Assume that Assumptions \ref{assum:1}, \ref{assum:2}, and \ref{assum:cz} hold.  Choose $\bar{p}=p+k$, $\bar{q}=q+k$, and $\bar{r}=r+k$ with $k$ being any positive integer satisfying $k>\frac{\log [\|C(z)\|^{-1}_{\infty} K^{-1} (1-\lambda)]}{|\log \lambda|}-1,$ where $\|C(z)\|^{2}_{\infty}=\max \limits_{|z|=1} \lambda_{\max} \{C(z)C^\top(z^{-1})\}$ and $K>0$ and $\lambda\in(0,1)$ are specified in Remark \ref{rm:cz}. For $\{\widehat{w}_N\}_{N\geq 1}$ generated by Algorithm \ref{Alg:1}, it holds that 
$
\sum_{k=1}^N \|\widehat{w}_{k+1}-w_{k+1}\|^2
=O\left(\log \lambda_{\max}(N)\right),~~{\rm a.s.}
$
\end{lemma}

%

\subsection{Consistency in Parameter Identification }\label{sec:alg1ttheorys1}

\begin{assumption}\label{assum:3}
For the maximal and minimal eigenvalues of $\mu I_d+\sum_{k=1}^{N} \varphi_k \varphi_k^\top$,
$
\lambda_{\min}(N) \mathop{\longrightarrow} \limits_{N \rightarrow \infty} \infty,$ $
\frac{\log\lambda_{\max}(N)}{\lambda_{\min}(N)} \mathop{\longrightarrow} \limits_{N \rightarrow \infty} 0$, $\mathrm{a.s.}
$
\end{assumption}

\begin{remark}\label{Rmk:3}
{\color{black}{For the classical persistent excitation (PE) condition, it usually requires that $\frac{\lambda_{\max}(N)}{\lambda_{\min}(N)}=O(1),~N\geq1$. It is direct to check that Assumption \ref{assum:3} allows that the sequence $\big\{\frac{\lambda_{\max}(N)}{\lambda_{\min}(N)}\big\}_{N\geq1}$ can be unbounded and includes the PE condition as a special case. In fact, by introducing the auxiliary variables, the alternating optimization strategy proposed in this paper renders the required excitation condition much weaker than those employed in existing literature. }}
{\color{black}{In Table \ref{Tab:0}, we compare the excitation conditions for sparse parameter identification investigated in literature, c.f., the algorithm in \cite{li2020online}, the algorithm in \cite{fu2023support,zhao2020sparse}, the adaptive LASSO algorithm in \cite{zou2006adaptive}, and Algorithm \ref{Alg:1} in this paper. It is direct to check that the excitation condition in this paper includes non-stationary observations and is much weaker than those in the above literature.}}
\begin{table*}[hbpt!]
{
\footnotesize
\caption{Excitation Conditions for Sparse Parameter Identification}
\label{Tab:0}
\begin{center}
\begin{tabular}{ccc}
\toprule
\bfseries &  \textbf{Type of Algorithms}  & \textbf{Excitation Conditions} \\
\hline
\textbf{Algorithm \ref{Alg:1}} & recursive & $\lambda_{\min }(N) \mathop{\longrightarrow} \limits_{N \rightarrow \infty} \infty,~
\frac{\log\lambda_{\max}(N)}{\lambda_{\min}(N)} \mathop{\longrightarrow} \limits_{N \rightarrow \infty} 0$   \\
\textbf{Algorithm in \cite{li2020online}} & recursive & $\liminf \limits_{N \rightarrow \infty} \frac{\lambda_{\min }(N)}{N^\alpha}>\frac{2\mu}{\rho}$, for some $\rho \in (0,1)$ and $\alpha>1$  \\
\textbf{Algorithm in \cite{fu2023support,zhao2020sparse}} & non-recursive & $\frac{\lambda_{\max}(N)}{\lambda_{\min}(N)}\sqrt{\frac{\log\lambda_{\max}(N)}{\lambda_{\min}(N)}} \mathop{\longrightarrow} \limits_{N \rightarrow \infty} 0$ \\
\textbf{Adaptive LASSO in \cite{zou2006adaptive}} & non-recursive & $\frac{1}{N} \sum_{k=1}^N \varphi_k \varphi_k^\top \mathop{\longrightarrow} \limits_{N \rightarrow \infty} C$ for some matrix $C>0$ \\
\bottomrule
\end{tabular}
\end{center}
}
\end{table*}
\end{remark}

\begin{assumption}\label{assum:recoe}
For the regularization coefficients $\{\lambda_N\}_{N\geq 1}$ in (\ref{eq:thetahat'}), 
$
\lambda_N\!=\!O \left(\sqrt{\frac{\log\lambda_{\max}(N)}{\lambda_{\min}(N)}}\right),\frac{\log\lambda_{\max}(N)}{\lambda_{\min}(N)}\!=\!o(\lambda_N),~~\mathrm{a.s.}
$
\end{assumption}

\begin{theorem}\label{th:thetan}
Assume that Assumptions \ref{assum:1}, \ref{assum:2}, \ref{assum:cz}, \ref{assum:3}, and \ref{assum:recoe} hold. Choose positive integers $\bar{p}$, $\bar{q}$, and $\bar{r}$ that satisfy the conditions of Lemma \ref{lem:noise}. Then for $\{\Theta_N\}_{N\geq1}$ generated by Algorithm \ref{Alg:1}, it holds that
\begin{align}
\|\Theta_{N+1}-\Theta\|=O\left(\sqrt{\frac{\log\lambda_{\max}(N)}{\lambda_{\min}(N)}}\right),~~\mathrm{a.s.}\label{eq24}
\end{align}
\end{theorem}

\begin{proof}
Results similar to (\ref{eq24}) also hold for the classical least squares (LS) algorithm \cite{chen2012identification} and the online algorithms in \cite{li2020online}. Due to the fact that algorithms (\ref{eq:thetanrls3})--(\ref{eq:thetanrls2}) are different from those in \cite{chen2012identification,li2020online}, here we need an essentially different analysis procedure. First, we prove that the estimates $\{\Theta_N\}_{N\geq1}$ are bounded. Second, we establish the convergence rate of $\{\|\Theta-\Theta_N\|\}_{N\geq1}$. As will be seen in the proof, the re-weighted scheme (\ref{eq:thetahat'}) for the $L_1$ penalty plays an important role. \textcolor{blue}{The detailed proof is given in the Appendix.} 
\end{proof}

\subsection{Accuracy in Sparsity Recovery }\label{sec:alg1ttheorys2}

\begin{theorem}\label{th:alg1step2}
Assume that Assumptions \ref{assum:1}, \ref{assum:2}, \ref{assum:cz}, \ref{assum:3}, and \ref{assum:recoe} hold. Choose positive integers $\bar{p}$, $\bar{q}$, and $\bar{r}$ that satisfy the conditions of Lemma \ref{lem:noise}. Then for $\{\Xi_N\}_{N\geq1}$ generated by Algorithm \ref{Alg:1},
\begin{align}\label{eq:xierr}
\left\|\Xi_{N+1}-\Theta\right\| =O\left(\sqrt{\frac{\log\lambda_{\max}(N)}{\lambda_{\min}(N)}}\right),~\mathrm{a.s.}
\end{align}
And there exists an $\omega$-set $\Omega_0$ with $\mathbb{P}\{\Omega_0\}=1$ such that for any $\omega \in \Omega_0$, there exists an integer $N_0(\omega)>0$ such that
\begin{align}\label{eq26}
\mathcal{A}\left(\Xi_{N+1}\right)=\mathcal{A}^*,~N \geq N_0(\omega),
\end{align}
where $\mathcal{A}(\Xi_{N+1})$ and $\mathcal{A}^*$ are defined by (\ref{eq:AN}) and (\ref{eq:A}), respectively.
\end{theorem}

\begin{proof}
First, we prove the consistency of $\{\Xi_N\}_{N\geq1}$, i.e., $\left\|\Xi_{N+1}-\Theta\right\| =O\left(\sqrt{\frac{\log\lambda_{\max}(N)}{\lambda_{\min}(N)}}\right),~\mathrm{a.s.}$ Then we establish the accuracy in sparse recovery, i.e., $\mathcal{A}(\Xi_{N+1})=\mathcal{A}^*$ for all $N$ large enough. The proof is motivated by \cite{zhao2020sparse}. Since in this paper the algorithms and the technical assumptions are different from \cite{zhao2020sparse}, the analysis procedure has little in common with  \cite{zhao2020sparse}. \textcolor{blue}{The detailed proof is given in the Appendix.}
\end{proof}

\begin{remark}
Using the same notations as adopted in this paper, in \cite{fu2023support,zhao2020sparse} it proves that for the estimate $\Xi_N$ for sparsity recovery (generated by non-recursive algorithms)
$$
\left\|\Xi_{N+1}\!-\!\Theta\right\| \!=\! O\left(\frac{\lambda_{\max}(N)}{\lambda_{\min}(N)}\sqrt{\frac{\log\lambda_{\max}(N)}{\lambda_{\min}(N)}} \!+\!\frac{\lambda_N}{\lambda_{\min}(N)}\right).
$$
In comparison with \cite{fu2023support,zhao2020sparse}, Theorem \ref{th:alg1step2} establishes stronger results under relaxed input-output observation conditions. Furthermore, the algorithms presented herein are inherently recursive. 
\end{remark}

By Theorems \ref{th:thetan} and \ref{th:alg1step2}, for $\{S_N\}_{N\geq1}$ generated from (\ref{eq21}), it directly follows that

\begin{corollary}\label{coro:alg1}
Assume that Assumptions \ref{assum:1}, \ref{assum:2}, \ref{assum:cz}, \ref{assum:3}, and \ref{assum:recoe} hold. Choose positive integers $\bar{p}$, $\bar{q}$, and $\bar{r}$ that satisfy the conditions of Lemma \ref{lem:noise}. Then for $\{S_N\}_{N\geq1}$ generated from (\ref{eq21}), 
$
\left\|S_{N+1}-\Theta\right\| =O\left(\sqrt{\frac{\log\lambda_{\max}(N)}{\lambda_{\min}(N)}}\right),~\mathrm{a.s.}
$ 
And there exists an $\omega$-set $\Omega_0$ with $\mathbb{P}\{\Omega_0\}=1$ such that for any $\omega \in \Omega_0$, there exists an integer $N_0(\omega)>0$ such that 
$
\mathcal{A}\left(S_{N+1}\right)=\mathcal{A}^*,~N \geq N_0(\omega),
$ 
where $\mathcal{A}(S_{N+1})$ and $\mathcal{A}^*$ are defined by (\ref{eq:AN}) and (\ref{eq:A}), respectively.
\end{corollary}

\section{Theoretical Properties of Recursive Algorithms with Conditions on ${\bf \{\varphi_k^0\}_{k=1}^N}$}\label{sec:alg1theory0}

Note that in Algorithm \ref{Alg:1} $\widehat{w}_k$ is the estimate of the system noise $w_k$. It is natural to ask whether Assumption \ref{assum:3} can be transformed into conditions directly imposed on $\sum^N_{k=1} \varphi^0_k\varphi^{0\top}_k$, since the latter would be easier to be verified in practice. We introduce the following assumptions.

\begin{assumption}\label{assum:3_0}
For the maximal and minimal eigenvalues of matrix $\mu I_d+\sum_{k=1}^{N} \varphi_k^0 \varphi_k^{0\top}$, i.e., $\lambda_{\max}^0(N)=\lambda_{\max}\left\{\sum_{k=1}^N \varphi^0_k\varphi_k^{0\top}\right\}$, $\lambda_{\min}^0(N)=\lambda_{\min}\left\{\sum_{k=1}^N \varphi^0_k\varphi_k^{0\top}\right\}$,
$
\lambda_{\min }^0(N) \mathop{\longrightarrow} \limits_{N \rightarrow \infty} \infty,$ $
\frac{\log\lambda_{\max}^0(N)}{\lambda_{\min}^0(N)} \mathop{\longrightarrow} \limits_{N \rightarrow \infty} 0,~~\mathrm{a.s.}
$
\end{assumption}
\begin{theorem}\label{th:thetan0}
Assume that Assumptions \ref{assum:1}, \ref{assum:2}, \ref{assum:cz}, \ref{assum:recoe}, and \ref{assum:3_0} hold. Choose positive integers $\bar{p}$, $\bar{q}$, and $\bar{r}$ that satisfy the conditions of Lemma \ref{lem:noise}. Then for $\{\Theta_N\}_{N\geq1}$ generated by Algorithm \ref{Alg:1}, it holds that
\begin{align}
\|\Theta_{N+1}-\Theta\|=O\left(\sqrt{\frac{\log\lambda_{\max}^0(N)}{\lambda_{\min}^0(N)}}\right),~~\mathrm{a.s.}\label{eq:thetaerr0}
\end{align}
\end{theorem}
\begin{proof}
    Due to space limitation, here we only list the sketch. By Assumption \ref{assum:3_0}, similar with the proof of (4.66) and (4.67) in \cite{chen2012identification}, we have $\lambda_{\max }(N)=O\left(\lambda_{\max}^0(N)\right)$ and $\lambda_{\min }^0(N)=O\left(\lambda_{\min } (N)\right)$ which indicates that Assumption \ref{assum:3} holds. Theorem \ref{th:thetan} ensures that (\ref{eq24}) holds, which in turn implies (\ref{eq:thetaerr0}). 
\end{proof}
\begin{theorem}\label{th:alg1step20}
Assume that Assumptions \ref{assum:1}, \ref{assum:2}, \ref{assum:cz}, \ref{assum:recoe}, and \ref{assum:3_0} hold. Choose positive integers $\bar{p}$, $\bar{q}$, and $\bar{r}$ that satisfy the conditions of Lemma \ref{lem:noise}. Then for $\{\Xi_N\}_{N\geq1}$ generated by Algorithm \ref{Alg:1}, 
$
\left\|\Xi_{N+1}-\Theta\right\| =O\left(\sqrt{\frac{\log\lambda_{\max}^0(N)}{\lambda_{\min}^0(N)}}\right),~\mathrm{a.s.}
$ 
And there exists an $\omega$-set $\Omega_0$ with $\mathbb{P}\{\Omega_0\}=1$ such that for any $\omega \in \Omega_0$, there exists an integer $N_0(\omega)>0$ such that 
$
\mathcal{A}\left(\Xi_{N+1}\right)=\mathcal{A}^*,~N \geq N_0(\omega),
$ 
where $\mathcal{A}(\Xi_{N+1})$ and $\mathcal{A}^*$ are defined by (\ref{eq:AN}) and (\ref{eq:A}), respectively.
\end{theorem}
\begin{proof}
Combining the definition of $\widehat{\Theta}_{N+1}(s,t)$ and Assumption \ref{assum:3_0}, similar with the proof of Theorem \ref{th:alg1step2}, the conclusion of Theorem \ref{th:alg1step20} can be demonstrated. 
\end{proof}
\begin{corollary}\label{coro:alg10}
Assume that Assumptions \ref{assum:1}, \ref{assum:2}, \ref{assum:cz}, \ref{assum:recoe}, and \ref{assum:3_0} hold. Choose positive integers $\bar{p}$, $\bar{q}$, and $\bar{r}$ that satisfy the conditions of Lemma \ref{lem:noise}. Then for $\{S_N\}_{N\geq1}$ generated from (\ref{eq21}), 
$
\left\|S_{N+1}-\Theta\right\| =O\left(\sqrt{\frac{\log\lambda_{\max}^0(N)}{\lambda_{\min}^0(N)}}\right),~\mathrm{a.s.}
$ 
And there exists an $\omega$-set $\Omega_0$ with $\mathbb{P}\{\Omega_0\}=1$ such that for any $\omega \in \Omega_0$, there exists an integer $N_0(\omega)>0$ such that 
$
\mathcal{A}\left(S_{N+1}\right)=\mathcal{A}^*,~N \geq N_0(\omega),
$ 
where $\mathcal{A}(S_{N+1})$ and $\mathcal{A}^*$ are defined by (\ref{eq:AN}) and (\ref{eq:A}), respectively.
\end{corollary}

\section{Numerical Examples}\label{sec:sim}

The simulation study is conducted on a MS laptop equipped with an Intel Core i7-13800H CPU (2.90GHz).

{\em Example 1)} Consider a multivariate ARMAX system 
$
y_{k+1} + A_1 y_{k} + A_2 y_{k-1} = B_1 u_k + B_2 u_{k-1} + w_{k+1} + C_1 w_{k} + C_2 w_{k-1}
$ 
where $u_k \in \mathbb{R}^{10}$, $y_k \in \mathbb{R}^{10}$ and $w_k \in \mathbb{R}^{10}$ are the input, the output, and the noise, respectively. Set the true parameter matrices as $A_1=-1\times I_{10}$, $A_2=0.5 \times I_{10}$, $B_1=I_{10}$, $B_2=0.5 \times I_{10}$, $C_1=0.8 \times I_{10}$, and $C_2=0$. It is direct to check that $C(z)\triangleq I_{10}+C_1 z+ C_2 z^2$ is stable. 

{\color{black}{The noise sequence $\{w_k\}_{k\geq 0}$ is assumed to be i.i.d. following a Gaussian distribution $\mathcal{N}(0,\sigma^2 )$. Meanwhile, the random sequence $\{v_k\}_{k\geq 0}$ is also i.i.d. with a Gaussian distribution $\mathcal{N}(0,I_{10})$ and is independent of $\{w_k\}_{k\geq 0}$. Based on $\{v_k\}_{k\geq 0}$, the system input $\{u_k\}_{k\geq0}$ is defined as follows: for the first component of $u_k$,
$u_{k+1}(1)=u_{k}(1)+v_{k+1}(1),$ 
for the $i$-th component where $i=2,\cdots,10$,
$u_{k+1}(i)=0.5u_{k}(i)+v_{k+1}(i).$ 
Set $\Theta=[A_1,A_2,B_1,B_2,C_1,C_2]^{\top} \in \mathbb{R}^{60\times 10}$ and $\varphi_k^0=[y_k^{\top},y_{k-1}^{\top},u_k^{\top},u_{k-1}^{\top},w_k^{\top},w_{k-1}^{\top}]^{\top}$. It can be verified that the sequence $\{u_k\}_{k\geq0}$ is not asymptotically stationary. Correspondingly, the input-output observations of the ARMAX system fail to satisfy the persistent excitation condition, specifically, 
the quantity 
$\lambda_{\max}\{\sum_{k=1}^N\varphi^0_k\varphi^0_k\}/\lambda_{\min}\{\sum_{k=1}^N\varphi^0_k\varphi^0_k\}\to\infty, ~~\mathrm{as}~N\to\infty.$ It can also be verified that Assumption \ref{assum:3_0} is satisfied by the input-output observations.}} 

In this example, we test the performance of the algorithm with different noise variances, i.e., $\sigma^2 = 0.25,~0.5,~1,$ and $2$, and for each noise variance we do $100$ simulations, i.e., $T=100$. {\color{blue}{For a rational matrix transfer function $H(z)$, its $\mathcal{H}_2$ norm is defined as 
$\|H(z)\|_2^2 \triangleq \frac{1}{2\pi} \int_{-\pi}^{\pi} \|H(e^{-j\omega})\|_F^2 \, \mathrm{d}\omega.$ Define $
y_{k+1}^{\mathrm{signal}} \triangleq A(z)^{-1}B(z)u_k$ and $y_{k+1}^{\mathrm{noise}} \triangleq A(z)^{-1}C(z)w_{k+1}.
$ The signal-to-noise ratio (SNR) for each entry of $y_{k+1}$ is given by $
{\rm SNR}_k(i) \triangleq \frac{{\rm Var}[y_{k}^{\rm signal}(i)]}{{\rm Var}[y_{k}^{\rm noise}(i)]},~~i = 1,\cdots,10.
$ It is direct to calculate that for channels $i=2,\ldots,10$, the resulting SNRs are $6.73$, $3.37$, $1.68$, and $0.84$, corresponding to $\sigma^2 = 0.25,~0.5,~1,$ and $2$, respectively. For channel $i=1$ at time instant $k=5000$, the corresponding SNRs are $2.7709\times 10^{+4}$, $1.3855\times 10^{+4}$, $6.9273\times 10^{+3}$, and $3.4637\times 10^{+3}$.}} To evaluate the performance of the algorithm, we consider the following three indexes: parameter estimation error (PEE), correct rate (CR) of sparsity recovery of $\Theta$, and calculation time (CT), i.e., 
$\textbf{PEE}(N)\triangleq \frac{1}{T}\sum_{i=1}^T \left\|X_{N,i}-\Theta\right\|/\left\|\Theta\right\|,$ where $X_{N,i}$ is the estimate generated during the $i$-th simulation with the data length $N$, 
$\textbf{CR}(N)
\triangleq \frac{1}{T}\sum_{i=1}^T \mathbb{I}\{|X_{N,i}(s,t)|<\tau,~\forall (s,t)\in \mathcal{A}^*\}
\cdot \mathbb{I}\{|X_{N,i}(s,t)|\geq\tau,~\forall (s,t)\in {\mathcal{A}^*}^c\},$ 
with the threshold $\tau = 10^{-7}$ and 
$
\textbf{CT}(N)
\triangleq \frac{1}{T}\sum_{i=1}^T \sum_{j=1}^{N/h} {\rm time}(X_{j h,i})
$ 
where ${\rm time}(X_{jh,i})$ is the calculation time of the estimate $X_{j h,i}$ during the $i$-th simulation with the data length $jh$. 


With the {\em a posterior} estimates of the system noise and the available data set $\{\varphi_k,y_{k+1}\}_{k=1}^N$, we can compare the performance of different classes of algorithms. We compare the performance of Algorithm \ref{Alg:1} with two classes of recursive/online algorithms, i.e., the recursive least squares (RLS) algorithm based on the over-parameterization technique \cite{chen2012identification} and the online alternating minimization (OAM) algorithm in \cite{li2020online}. The tuning parameters for each algorithm are selected via cross-validation: for Algorithm \ref{Alg:1}, $\bar{p}=\bar{q}=\bar{r}=4$, $\mu=1$ and $\lambda_N=\big({\log N}/{N}\big)^{0.5}$, for RLS, the tuning parameter $\mu=10000$, while for OAM, $\mu=1$ and $\lambda=0.1$. 
We also compare Algorithm \ref{Alg:1} with two classes of offline algorithms, i.e., the least squares algorithm with weighted $L_1$ regularization (LSW, \cite{fu2023support,zhao2020sparse}) with the tuning parameter $\lambda_N=N^{0.8}$ and the SINDy algorithm \cite{brunton2016discovering} with the corresponding tuning parameter $\lambda=0.1$.


\begin{figure}[hbpt!]
\centering
\includegraphics[width=0.4\textwidth]{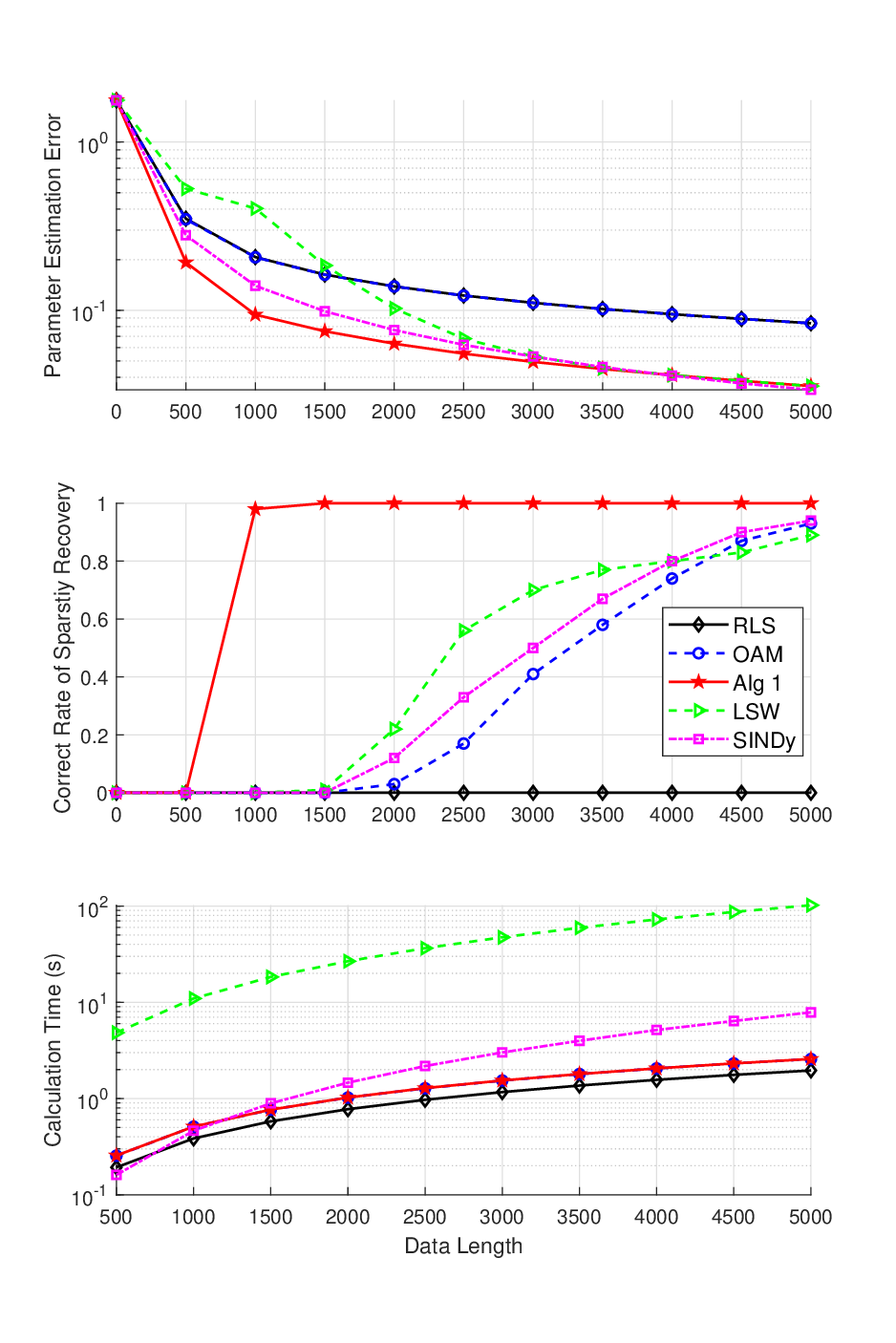}
\caption{Parameter estimation errors, correct rates of sparsity recovery, and calculation time of estimates generated by Algorithm \ref{Alg:1}, RLS, OAM, LSW, and SINDy with $\sigma^2=0.5$.}
\label{Fig:re1}
\end{figure}

\begin{table}[hbpt!]
{
\caption{PEE and CR for \(N=5000\) over 100 independent simulations.}
\label{Tab:re1}
\begin{center}
\begin{tabular}{ccccc}
\toprule
& \multicolumn{4}{c}{$\sigma^2$} \\
\cmidrule(lr){2-5}
\bfseries \textbf{PEE} & $0.25$ &  $0.5$ & $1$ & $2$\\
\hline
RLS & 0.0547 & 0.0638 & 0.0794 & 0.1022\\
OAM & 0.0546 & 0.0638 & 0.0794 & 0.1022\\
Algorithm \ref{Alg:1} & {\bf 0.0137} & {\bf 0.0199} & {\bf 0.0308} & {\bf 0.0481}\\
LSW & 0.0138 & 0.0200 & 0.0311 & 0.0488\\
SINDy & 0.0160 & 0.0225 & 0.0317 & 0.0450
\\
\bottomrule
& \multicolumn{4}{c}{$\sigma^2$} \\
\cmidrule(lr){2-5}
\textbf{CR} & $0.25$ &  $0.5$ & $1$ & $2$\\
\hline
RLS & 0 & 0 & 0 & 0\\
OAM & {\bf 1} & 0.98 & 0.96 & 0.64\\
Algorithm \ref{Alg:1} & {\bf 1} & {\bf 1} & {\bf 1} & {\bf 1}\\
LSW & 0.99 & 0.74 & 0.48 & 0.18\\
SINDy & {\bf 1} & 0.98 & 0.96 & 0.77\\
\bottomrule
\end{tabular}
\end{center}
}
\end{table}


Note that the conventional RLS algorithm cannot exploit the sparsity structure of the parameter matrix. 
Figure \ref{Fig:re1} and Table \ref{Tab:re1} show that, Algorithm \ref{Alg:1} achieves more accurate sparsity recovery and a smaller parameter estimation error than the RLS, OAM, LSW and SINDy algorithms, and exhibits higher computational efficiency than the offline sparsity recovery algorithms (e.g., LSW and SINDy). 

{\em Example 2)} Consider a high-dimensional linear stochastic system $y_{k}=\Theta^\top \varphi_k+w_k$, $k\geq 1$ where $\varphi_k\in \mathbb{R}^{d}$, $y_k \in \mathbb{R}^{n}$ and $w_k \in \mathbb{R}^{n}$ are the regressor, the output and the noise, respectively. 
Set $d=n=100$ and the number of Monte Carlo simulations $T=100$. In each simulation trial, the parameter matrix $\Theta\in \mathbb{R}^{100 \times 100}$ is constructed with $25\%$ nonzero entries. The positions of these nonzero entries are randomly selected, and all nonzero entries are fixed to unity. 
For $k \geq 1$, set $\varphi_k$ as follows: for $i = 1, \dots, 50$, $\varphi_{k+1}(i) = \varphi_k(i) + v_{k+1}(i)$; for $j = 51, \dots, 100$, $\varphi_{k+1}(j) = 0.5 \varphi_k(j) + v_{k+1}(j)$, where $\{v_k\}_{k\geq1}$ is i.i.d. with a Gaussian distribution $\mathcal{N}(0,I_{d})$. It is direct to verify that $\{\varphi_k\}_{k\geq 1}$ is nonstationary and does not satisfy the PE condition.  
The noise $\{w_k\}_{k\geq 1}$ is assumed to i.i.d. with a Gaussian distribution $\mathcal{N}(0,\sigma^2 )$ which is independent of $\{\varphi_k\}_{k\geq 1}$. With the maximum data length of $N=5000$, we compare the performance of Algorithm 1 with that of the RLS, OAM, LSW, and SINDy algorithms, where the hyperparameters are identical to those adopted in Example 1).

\begin{figure}[hbpt!]
\centering
\includegraphics[width=0.4\textwidth]{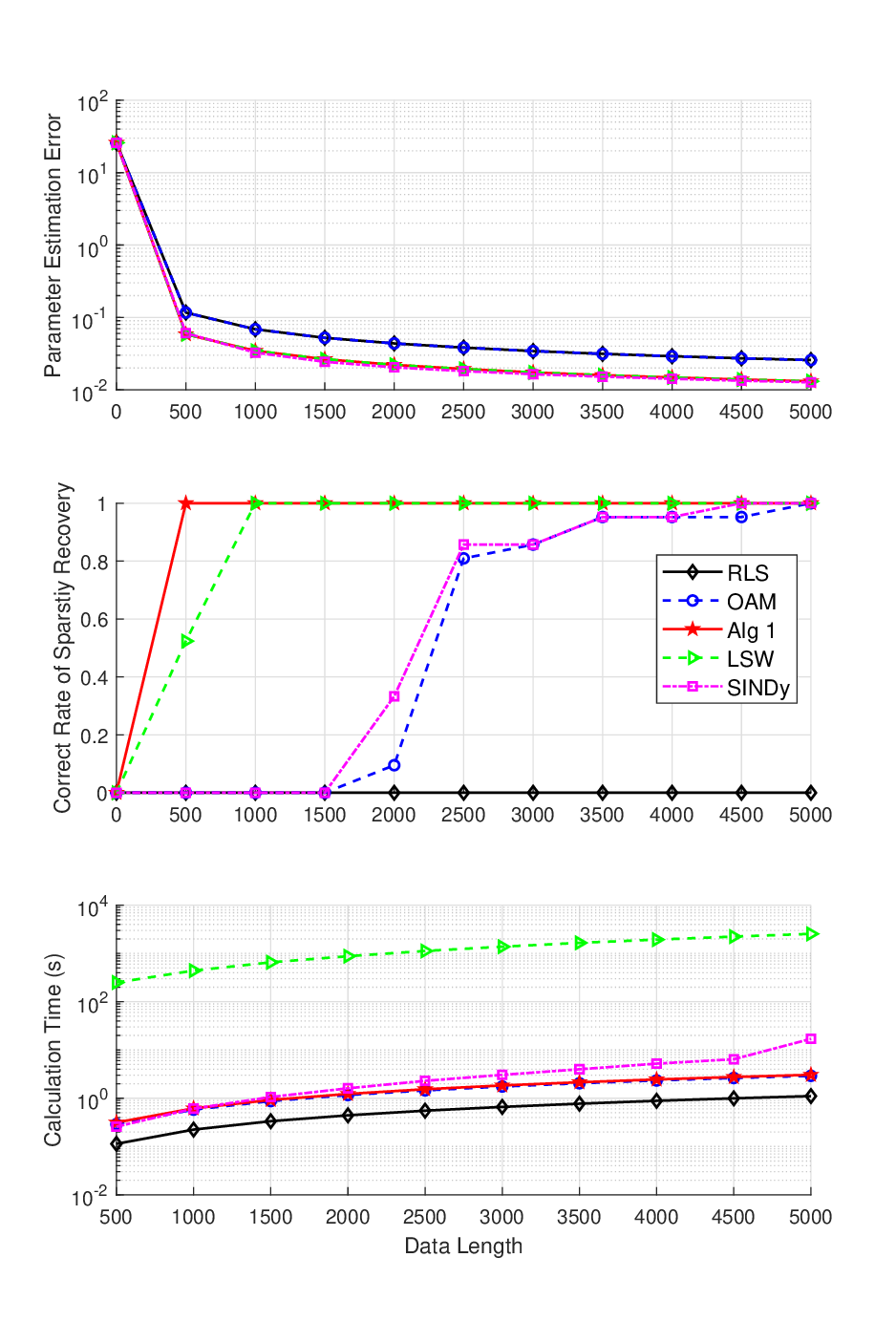}
\caption{Parameter estimation errors, correct rates,  and calculation time of estimates generated by Algorithm 1, RLS, OAM, LSW, and SINDy with $\sigma^2=2$.}
\label{Fig:re3}
\end{figure}

Figure \ref{Fig:re3} and Table \ref{Tab:re3} illustrate the performance of Algorithm 1, RLS, OAM, LSW, and SINDy for $\sigma^2=2$. Figure \ref{Fig:re3} and Table \ref{Tab:re3} show that, Algorithm \ref{Alg:1} achieves higher accuracy in sparsity recovery than the RLS, OAM, LSW and SINDy algorithms with a small number of observations, and exhibits higher computational efficiency than the offline sparsity recovery algorithms (e.g., LSW and SINDy). 

\begin{table}[hbpt!]
{
\caption{Mean computational time (CT) (in seconds) for \(N=5000\) over 100 independent simulations.}
\label{Tab:re3}
\begin{center}
\begin{tabular}{ccccc}
\toprule
 RLS & OAM & Algorithm \ref{Alg:1} & LSW & SINDy \\
\hline
1.1076 & 2.9105 & 3.0589 & 2547.4100 & 17.0084\\
\bottomrule
\end{tabular}
\end{center}
}
\end{table}

\section{Conclusion}

This paper proposes recursive algorithms for sparse parameter identification of multivariate ARMAX systems with non-stationary observations and colored noise and theoretical properties of the recursive algorithms are established, that is, the accurate estimation of the sparse index set of the unknown parameter matrix and the consistent estimates for values of the non-zero entries of the unknown parameter matrix can be achieved simultaneously. {\color{blue}{Note that the above estimation error bounds are sample path-dependent. Establishing explicit $N$-dependent probability bounds on the estimation errors of $\|\Theta_{N+1}-\Theta\|$ and $\|\Xi_{N+1}-\Theta\|$ would be an interesting topic for future research.}} Other future work could explore the derivation of recursive solutions for non-convex penalties (e.g., \cite{Ghayem2018Sparse}) in sparsity recovery, etc. 


\section*{Appendix}


The following lemma is well established in literature.

\begin{lemma}(\cite{mds1986})\label{lem:mds}
Assume that Assumptions \ref{assum:1} and \ref{assum:2} hold. Then
\begin{align}
&\left\|\left(\sum\nolimits_{k=1}^{N} \varphi_k \varphi_k^{\top}\right)^{-1/2}\sum\nolimits_{k=1}^{N} \varphi_k w_{k+1}^{\top}\right\|\nonumber\\
&=O\left(\sqrt{\log \lambda_{\max}\left\{\sum\nolimits_{k=1}^{N} \varphi_k \varphi_k^{\top}\right\}}\right)~~\mathrm{a.s.}
\end{align}
\end{lemma}

\subsection*{Proof of Theorem \ref{th:thetan}}\label{app:th1}

Noting that Assumption \ref{assum:3} holds almost surely, there exists an $\omega$-set $\Omega_{0}$ with $\mathbb{P}\{\Omega_{0}\}=1$ such that Assumption \ref{assum:3} holds for any $\omega \in \Omega_{0}$. \textcolor{black}{In the following proof,} we will analyze the sequence of estimates on a fixed sample path $\omega\in\Omega_0$.

Denote $\widetilde{\Theta}_{N+1}\triangleq \Theta-\Theta_{N+1}$,  $\widetilde{\Xi}_{N+1}\triangleq \Theta-\Xi_{N+1}$ and $\varphi^\xi_N \triangleq \varphi_N-\varphi_N^0$. 
\textcolor{black}{For the sub-optimization problem (\ref{pro0:theta}), it has a straightforward closed-form solution 
\begin{align}
\Theta_{N+1}
=\left(\mu I_d+\Phi_N^\top \Phi_N\right)^{-1}\left(\Phi_N^\top Y_{N+1}+\mu \Xi_{N}\right)\label{eq:thetan}.
\end{align}
Recalling $P_{N+1}=(\mu I_d+\sum_{k=1}^N \varphi_k \varphi_k^\top)^{-1}$, we obtain that
\begin{align}\label{eq:tildetheta}
&\widetilde{\Theta}_{N+1}
=\Theta-P_{N+1}\sum_{k=1}^{N} \varphi_k y_{k+1}^\top
-\mu P_{N+1} \Xi_{N}\nonumber\\
&=\Theta-P_{N+1} \sum_{k=1}^{N}\varphi_k\left(\varphi_k^{0\top} \Theta + w_{k+1}^\top\right)
-\mu P_{N+1} \Xi_{N}\nonumber\\
&=P_{N+1}\Big(\mu I_d+\sum_{k=1}^N \varphi_k \varphi_k^\top\Big)\Theta-P_{N+1} \sum_{k=1}^{N}\varphi_k\varphi_k^\top \Theta\nonumber\\
&~~~~+P_{N+1} \sum_{k=1}^{N}\varphi_k\varphi_k^{\xi\top} \Theta
-P_{N+1} \sum_{k=1}^{N}\varphi_k w_{k+1}^\top-\mu P_{N+1} \Xi_{N}\nonumber\\
&=\mu P_{N+1} \widetilde{\Xi}_{N}+P_{N+1} \sum_{k=1}^{N}\varphi_k\varphi_k^{\xi\top} \Theta - P_{N+1} \sum_{k=1}^{N}\varphi_k w_{k+1}^\top.
\end{align}
}

Noting (\ref{eq:sparseso}), it follows that
\begin{align}
&\Xi_{N+1}(s,t)=\nonumber\\
&\left\{
\begin{matrix}
{\rm sgn}\left(\Theta_{N+1}(s,t)\right) \Big(\left|\Theta_{N+1}(s,t)\right|-\frac{\lambda_N}{\mu|\widehat{\Theta}_{N+1}(s,t)|}\Big), \\
\text{~~~~~if } \left|\Theta_{N+1}(s,t)\right|\geq \frac{\lambda_N}{\mu |\widehat{\Theta}_{N+1}(s,t)|}, \\
0, \text{~~if } \left|\Theta_{N+1}(s,t)\right|< \frac{\lambda_N}{\mu |\widehat{\Theta}_{N+1}(s,t)|}.
\end{matrix}
\right.\label{eq:XiNdef}
\end{align}

By the definition of $\widehat{\Theta}_{N+1}(s,t)$ given in (\ref{eq:thetahat}), we can get
\begin{align}\label{eq:thetahatabs}
\left|\widehat{\Theta}_{N+1}(s,t)\right|
&=\left|\Theta_{N+1}(s,t)\right|+\sqrt{\frac{\log\lambda_{\max}(N)}{\lambda_{\min}(N)}}\nonumber\\
&\geq \sqrt{\frac{\log\lambda_{\max}(N)}{\lambda_{\min}(N)}}.
\end{align}

Combining (\ref{eq:XiNdef}) and (\ref{eq:thetahatabs}), for any $N\geq 1$, we further obtain
\begin{align}
|\Xi_{N+1}(s,t)-\Theta_{N+1}(s,t)|
&\leq \frac{\lambda_N}{\mu \big|\widehat{\Theta}_{N+1}(s,t)\big|}\nonumber\\
&\leq \frac{\lambda_N}{\mu \sqrt{\frac{\log \lambda_{\max}(N)}{\lambda_{\min}(N)}}}.\label{eq71}
\end{align}

Define $\beta_{N+1}\triangleq \widetilde{\Xi}_{N+1}-\widetilde{\Theta}_{N+1}$. We have that
\begin{align}\label{eq:xitheta}
\widetilde{\Xi}_{N+1}=\widetilde{\Theta}_{N+1}+\beta_{N+1}
\end{align}
and by noting (\ref{eq71}),
\begin{align}\label{eq:beta}
\|\beta_{N+1}\|\leq \frac{\sqrt{dn}}{\mu} \frac{\lambda_N}{ \sqrt{\frac{\log \lambda_{\max}(N)}{\lambda_{\min}(N)}}}.
\end{align}

\textcolor{black}{
By substituting (\ref{eq:xitheta}) into (\ref{eq:tildetheta}) it yields
\begin{align}\label{eq:tildetheta0}
&\widetilde{\Theta}_{N+1}
=\mu P_{N+1} \widetilde{\Theta}_{N} + \mu P_{N+1} \beta_N \nonumber\\
&~~~~~~~~~~+P_{N+1} \sum_{k=1}^{N}\varphi_k\varphi_k^{\xi\top} \Theta
-P_{N+1} \sum_{k=1}^{N}\varphi_k w_{k+1}^\top.
\end{align}
For the third term in (\ref{eq:tildetheta0}), by Cauchy–Schwarz inequality, it follows that 
\begin{align}\label{eq:phixi0}
&\Big\|P_{N+1} \sum_{k=1}^{N}\varphi_k \varphi_k^{\xi \top} \Theta\Big\|
\leq \sum_{k=1}^{N} \|P_{N+1} \varphi_k\| \cdot  \|\varphi_k^{\xi \top} \Theta\|\nonumber\\
&\leq \Big(\sum_{k=1}^{N} \|P_{N+1}\varphi_k\|^2\Big)^{1/2} \cdot \Big(\sum_{k=1}^{N} \|\varphi_k^{\xi \top}\Theta\|^2\Big)^{1/2}
\end{align}
and 
\begin{align}\label{eq:phixi01}
    &\sum_{k=1}^{N} \|P_{N+1}\varphi_k\|^2 
    \leq \lambda_{\min}^{-1}(N)\sum_{k=1}^{N} \varphi_k^\top P_{N+1} \varphi_k \nonumber\\
    &= \lambda_{\min}^{-1}(N) {\rm tr}\Big(P_{N+1} \sum_{k=1}^{N} \varphi_k \varphi_k^\top  \Big)
    =O(\lambda_{\min}^{-1}(N)).
\end{align}
Noting $\varphi_k^\xi =[0_{1\times (np+lq)}, \widehat{w}_{k}^\top-w_k^\top,\cdots,\widehat{w}_{k-r+1}^\top-{w}_{k-r+1}^\top]^\top$, by Lemma \ref{lem:noise}, we have  
\begin{align}\label{eq:phixi02}
    \sum_{k=1}^{N} \|\varphi_k^{\xi \top}\Theta\|^2
    &=O\Big(\sum_{k=1}^{N} \|\widehat{w}_{k+1}-w_{k+1}\|^2\Big)\nonumber\\
    &=O\left(\log \lambda_{\max}(N)\right)
\end{align}
Substituting (\ref{eq:phixi01}) and (\ref{eq:phixi02}) into (\ref{eq:phixi0}), we can get
\begin{align}\label{eq:phixi}
    \Big\|P_{N+1} \sum_{k=1}^{N}\varphi_k \varphi_k^{\xi \top} \Theta\Big\|
    =O\left(\sqrt{\frac{\log \lambda_{\max}(N)}{\lambda_{\min}(N)}}\right).
\end{align}
}

For the fourth term in (\ref{eq:tildetheta0}), it yields that
\begin{align}\label{eq:mds0}
&\Big\|P_{N+1}^{1/2} \sum\nolimits_{k=1}^{N}\varphi_k w_{k+1}^\top\Big\|\nonumber\\
&\leq \Big\|P_{N+1}^{1/2}\Big(\sum_{k=1}^{N} \varphi_k \varphi_k^{\top}\Big)^{1/2}\Big\| \Big\| \Big(\sum_{k=1}^{N} \varphi_k \varphi_k^{\top}\Big)^{-1/2}\! \sum_{k=1}^{N}\! \varphi_k w_{k+1}^{\top}\Big\|.
\end{align}

Noting Assumption \ref{assum:3} that $\lambda_{\min }(N) \rightarrow \infty$ as $N \rightarrow \infty$, we have
$
\big\|\big(\sum_{k=1}^{N} \varphi_k \varphi_k^\top\big)^{-1}\big\|
\leq \sqrt{d} \lambda_{\min}^{-1}\big\{\sum_{k=1}^{N} \varphi_k \varphi_k^\top\big\} \mathop{\longrightarrow} \limits_{N \rightarrow \infty} 0
$
which implies $\big(\sum_{k=1}^{N} \varphi_k \varphi_k^\top\big)^{-1}\mathop{\longrightarrow} \limits_{N \rightarrow \infty} 0$. 
By the definition of $P_{N+1}$, it yields that
$
P_{N+1} \sum_{k=1}^{N} \varphi_k \varphi_k^\top
=\big(\mu I_d+\sum_{k=1}^{N} \varphi_k \varphi_k^\top\big)^{-1} \sum_{k=1}^{N} \varphi_k \varphi_k^\top
=\big(\mu \big(\sum\nolimits_{k=1}^{N} \varphi_k \varphi_k^\top\big)^{-1}+I_d\big)^{-1}
\mathop{\longrightarrow} \limits_{N \rightarrow \infty} I_d
$
and thus
\begin{align}\label{eq:lmaxp}
\Big\|P_{N+1}^{1/2}\Big(\sum\nolimits_{k=1}^{N} \varphi_k \varphi_k^{\top}\Big)^{1/2}\Big\|=O(1).
\end{align}

Substituting (\ref{eq:lmaxp}) into (\ref{eq:mds0}) and by Lemma \ref{lem:mds}, it follows that $\big\|P_{N+1}^{1/2} \sum_{k=1}^{N}\varphi_k w_{k+1}^\top\big\|=O\big(\sqrt{\log \lambda_{\max}(N)}\big)$ and for the last term in (\ref{eq:tildetheta0}),
\begin{align}\label{eq:noise}
\Big\|P_{N+1} \!\!\sum\nolimits_{k=1}^{N}\!\! \varphi_k w_{k+1}^\top\Big\|
&\leq\!\! \left\|P_{N+1}^{1/2}\right\| \Big\|P_{N+1}^{1/2} \!\!\sum\nolimits_{k=1}^{N}\!\!\varphi_k w_{k+1}^\top\Big\|\nonumber\\
&=O\left(\sqrt{\frac{\log \lambda_{\max}(N)}{\lambda_{\min}(N)}}\right).
\end{align}

\textcolor{black}{Denote $\xi_N \!\triangleq \!\big\|P_{N+1} \sum_{k=1}^{N}\varphi_k \varphi_k^{\xi \top} \Theta\big\|\!+\!\big\|P_{N+1} \sum_{k=1}^{N}\varphi_k w_{k+1}^\top\big\|$. Then by (\ref{eq:phixi}) and (\ref{eq:noise}) we have $\xi_N=O\left(\sqrt{\frac{\log \lambda_{\max}(N)}{\lambda_{\min}(N)}}\right)$.} By (\ref{eq:beta}) and (\ref{eq:tildetheta0}), for any $N\geq 1$ there exists a constant $c>0$ such that
\begin{align}\label{eq:thetat}
&\big\|\widetilde{\Theta}_{N+1}\big\|
\leq \big\|\mu P_{N+1} \widetilde{\Theta}_{N}\big\| + \left\|\mu P_{N+1} \beta_N\right\| +\xi_N\nonumber\\
&\leq\mu \lambda_{\min}^{-1}(N)\big\|\widetilde{\Theta}_{N}\big\|\!+\!\sqrt{dn} \lambda_{\min}^{-1}(N)\frac{\lambda_{N-1}}{\sqrt{\frac{\log \lambda_{\max}(N-1)}{\lambda_{\min}(N-1)}}}\!+\!\xi_N\nonumber\\
&\leq\mu \lambda_{\min}^{-1}(N)\big\|\widetilde{\Theta}_{N}\big\|+c \lambda_{\min}^{-1}(N)+\xi_N
\end{align}
where for the last inequality Assumption \ref{assum:recoe} is applied.

\textcolor{black}{By (\ref{eq:noise}) and Assumption \ref{assum:3},} we know that the positive sequence $\{\xi_N\}_{N\geq 1}$ is bounded and there exists a constant $c_0>0$ such that $\xi_N\leq c_0$ for any $N\geq 1$. From the inequality (\ref{eq:thetat}), we further obtain
\begin{align}\label{eq:thetat0}
&\big\|\widetilde{\Theta}_{N+1}\big\|
\leq\mu \lambda_{\min}^{-1}(N)\Big(\mu \lambda_{\min}^{-1}(N-1)\big\|\widetilde{\Theta}_{N-1}\big\|\nonumber\\
&+c \lambda_{\min}^{-1}(N-1)+\xi_{N-1}\Big)
+c \lambda_{\min}^{-1}(N)+\xi_{N}\nonumber\\
&\leq\mu^2 \lambda_{\min}^{-1}(N) \lambda_{\min}^{-1}(N-1)\big\|\widetilde{\Theta}_{N-1}\big\|\nonumber\\
&+ c\left( \mu \lambda_{\min}^{-1}(N) \lambda_{\min}^{-1}(N-1)+\lambda_{\min}^{-1}(N)\right)\nonumber\\
&+\mu \lambda_{\min}^{-1}(N) \xi_{N-1}+\xi_{N}\nonumber\\
&\leq\mu^2 \lambda_{\min}^{-1}(N) \lambda_{\min}^{-1}(N-1)\Big(\mu \lambda_{\min}^{-1}(N-2)\big\|\widetilde{\Theta}_{N-2}\big\|\nonumber\\
&+c \lambda_{\min}^{-1}(N-2)+\xi_{N-2}\Big) + c\Big( \mu \lambda_{\min}^{-1}(N) \lambda_{\min}^{-1}(N-1)\nonumber\\
&+\lambda_{\min}^{-1}(N)\Big)+\mu \lambda_{\min}^{-1}(N) \xi_{N-1}+\xi_{N}\nonumber\\
&\leq \mu^3 \lambda_{\min}^{-1}(N) \lambda_{\min}^{-1}(N-1)\lambda_{\min}^{-1}(N-2)\big\|\widetilde{\Theta}_{N-2}\big\| \nonumber\\
&+ \frac{c}{\mu}\Big( \mu^3 \lambda_{\min}^{-1}(N) \lambda_{\min}^{-1}(N-1)\lambda_{\min}^{-1}(N-2)\nonumber\\
&+\mu^2 \lambda_{\min}^{-1}(N) \lambda_{\min}^{-1}(N-1)+\mu\lambda_{\min}^{-1}(N)\Big)\nonumber\\
&+\mu^2 \lambda_{\min}^{-1}(N) \lambda_{\min}^{-1}(N-1)\xi_{N-2}
+\mu \lambda_{\min}^{-1}(N) \xi_{N-1}+\xi_{N}.
\end{align}
\textcolor{black}{Following a derivation analogous to (\ref{eq:thetat0}), we have}
\begin{align}\label{eq:thetat1}
&\big\|\widetilde{\Theta}_{N+1}\big\|\leq \mu^{N+1}\prod\nolimits_{k=0}^{N}\lambda_{\min}^{-1}(N-k)\big\|\widetilde{\Theta}_0\big\|\nonumber\\
&~~~~+\frac{c}{\mu}\left(\sum\nolimits_{k=1}^{N+1} \mu^k \prod\nolimits_{i=N-k+1}^{N}\lambda_{\min}^{-1}(i)\right)\nonumber\\
&~~~~+\left( \sum\nolimits_{k=1}^N\!\! \mu^k \left(\prod\nolimits_{i=N-k+1}^{N}\!\! \lambda_{\min}^{-1}(i)\right) \xi_{N-k}\right)\!+\!\xi_{N}\nonumber\\
&\leq \mu^{N+1}\prod\nolimits_{k=0}^{N}\lambda_{\min}^{-1}(N-k)\big\|\widetilde{\Theta}_0\big\|\nonumber\\
&~~~~+\frac{c}{\mu}\left(\sum\nolimits_{k=1}^{N+1} \mu^k \prod\nolimits_{i=N-k+1}^{N}\lambda_{\min}^{-1}(i)\right)\nonumber\\
&~~~~+c_0 \left( \sum\nolimits_{k=1}^N \mu^k \prod\nolimits_{i=N-k+1}^{N}\lambda_{\min}^{-1}(i) \right)+c_0
\end{align}
\textcolor{black}{where for the last inequality the boundedness of $\left\{\xi_N\right\}_{N\geq 1}$ is used.}

\textcolor{black}{From the inequality (\ref{eq:thetat1}),} it leads to
\begin{align}\label{eq:thetat2}
&\left\|\widetilde{\Theta}_{N+1}\right\|
= O\left(\mu^{N+1}\prod_{k=0}^{N}\lambda_{\min}^{-1}(N-k)\right)\nonumber\\
&+O\left(\sum_{k=1}^{N+1} \mu^k \prod_{i=N-k+1}^{N}\lambda_{\min}^{-1}(i)\right)+O(1).
\end{align}

Note that $\lambda_{\min}(0)=\mu$. For $N\geq 1$, we can get
$
\lambda_{\min}(N)
=\mu + \lambda_{\min}\big\{\sum_{k=1}^{N} \varphi_k \varphi_k^\top\big\}
\geq \mu
$
and thus
\begin{align}\label{eq:mumin}
\frac{\mu}{\lambda_{\min}(N)} \leq 1,~\forall N\geq 1.
\end{align}

For the first item in (\ref{eq:thetat2}), we have
\begin{align}\label{eq:term1}
\mu^{N+1}\prod_{k=0}^{N}\lambda_{\min}^{-1}(N-k)
=\prod_{k=0}^{N}\frac{\mu}{\lambda_{\min}(N-k)}\leq 1.
\end{align}

Noting Assumption \ref{assum:3} that $\lambda_{\min }(N) \mathop{\longrightarrow} \limits_{N \rightarrow \infty} \infty$, \textcolor{black}{for any fixed $\alpha\in (0,1)$,} there exists an integer $N_0>0$ large enough such that for any integer $N\geq N_0$,
$
\lambda_{\min}(N)\geq \frac{\mu}{\alpha}
$
which implies
\begin{align}\label{eq:mu}
\frac{\mu}{\lambda_{\min}(N)} \leq \alpha,~\forall N\geq N_0.
\end{align}

Hence for any integer $N\geq 2N_0-1$, which implies $N-N_0+1\geq N_0$, we have
\begin{align}\label{eq:Sn}
&\sum_{k=1}^{N+1} \mu^k \prod_{i=N-k+1}^{N}\lambda_{\min}^{-1}(i)\nonumber\\
&=\sum_{k=1}^{N_0} \mu^k \prod_{i=N-k+1}^{N}\lambda_{\min}^{-1}(i)+\sum_{k=N_0+1}^{N+1} \mu^k \prod_{i=N-k+1}^{N}\lambda_{\min}^{-1}(i)\nonumber\\
&=\sum_{k=1}^{N_0} \prod_{i=N-k+1}^{N} \frac{\mu}{\lambda_{\min}(i)}\nonumber\\
&+\sum_{k=N_0+1}^{N+1} \left(\prod_{i=N-k+1+N_0}^{N} \frac{\mu}{\lambda_{\min}(i)}\right)\left( \prod_{i=N-k+1}^{N-k+N_0} \frac{\mu}{\lambda_{\min}(i)}\right) \nonumber\\
&\leq N_0+\sum_{k=N_0+1}^{N+1} \left(\prod_{i=N-k+1+N_0}^{N} \frac{\mu}{\lambda_{\min}(i)}\right),
\end{align}
where for the inequality (\ref{eq:mumin}) is used.

By (\ref{eq:mu}) and noting $\alpha\in (0,1)$, we get
\begin{align}
&\sum_{k=N_0+1}^{N+1} \prod_{i=N-k+1+N_0}^{N} \frac{\mu}{\lambda_{\min}(i)}
\leq \sum_{k=N_0+1}^{N+1} \alpha^{k-N_0}\nonumber\\
&=\alpha^{-N_0} \sum_{k=N_0+1}^{N+1} \alpha^{k}<\infty.\label{eq89}
\end{align}
By (\ref{eq:Sn}) and (\ref{eq89}), it follows that
\begin{align}\label{eq:term2}
\limsup \limits_{N\rightarrow \infty} \sum_{k=1}^{N+1} \mu^k \prod_{i=N-k+1}^{N}\lambda_{\min}^{-1}(i)<\infty.
\end{align}

\textcolor{black}{Substituting (\ref{eq:term1}) and  (\ref{eq:term2}) into (\ref{eq:thetat2}),} we prove that $\{\|\widetilde{\Theta}_{N}\|\}_{N\geq1}$ is bounded,
\begin{align}
\limsup \limits_{N\rightarrow \infty} \|\widetilde{\Theta}_{N}\|<\infty.\label{eq91}
\end{align}

Noting (\ref{eq91}), there exists an integer $N_1>0$ large enough and a constant $c>0$ such that
$
\big\|\widetilde{\Theta}_{N}\big\|\leq c,~\forall N\geq N_1,
$
from which and also noting (\ref{eq:phixi}), (\ref{eq:noise}) and (\ref{eq:thetat}),
\begin{align}
\left\|\widetilde{\Theta}_{N+1}\right\|
&= O\left(\lambda_{\min}^{-1}(N)+\sqrt{\frac{\log \lambda_{\max}(N)}{\lambda_{\min}(N)}}\right)\nonumber\\
&= O\Bigg(\frac{1}{\sqrt{\lambda_{\min}(N)\log \lambda_{\max}(N)}} \sqrt{\frac{\log \lambda_{\max}(N)}{\lambda_{\min}(N)}}\nonumber\\
&~~~+\sqrt{\frac{\log \lambda_{\max}(N)}{\lambda_{\min}(N)}}\Bigg).\label{eq93}
\end{align}

Since $\lambda_{\min}(N) \rightarrow \infty$ as $N \rightarrow \infty$, it follows that
$
\frac{1}{\sqrt{\lambda_{\min}(N)\log \lambda_{\max}(N)}}\mathop{\longrightarrow}\limits_{N\to\infty}0
$
from which and (\ref{eq93}),
$
\big\|\widetilde{\Theta}_{N+1}\big\|
=O\left(\sqrt{\frac{\log \lambda_{\max}(N)}{\lambda_{\min}(N)}}\right).
$

This finishes the proof.

\subsection*{Proof of Theorem \ref{th:alg1step2}}

Note that $\Xi_{N+1}$ is the optimal point of (\ref{pro0:xi}). Denote
\begin{align}
F_{N+1}(T) \triangleq \frac{\mu}{2}\left\|\Theta_{N+1}-T\right\|^2
+\lambda_N \sum_{s=1}^{d} \!\sum_{t=1}^{n} \!\frac{\left|T(s,t)\right|}{\big|\widehat{\Theta}_{N+1}(s,t)\big|}.\nonumber
\end{align}

Recall the definition $\widetilde{\Xi}_{N+1}\triangleq \Theta-\Xi_{N+1}$. By the optimality of $\Xi_{N+1}$, we have
\begin{align}\label{eq:dF}
0&\geq F_{N+1}\left(\Xi_{N+1}\right)-F_{N+1}(\Theta)\nonumber\\
&=\frac{\mu}{2}\big\|\Theta_{N+1}-\big(\Theta-\widetilde{\Xi}_{N+1}\big)\big\|^2\nonumber\\
&+ \lambda_N \sum_{s=1}^{d} \sum_{t=1}^{n} \frac{1}{\big|\widehat{\Theta}_{N+1}(s,t)\big|}\big|\Theta(s,t)-\widetilde{\Xi}_{N+1}(s,t)\big|\nonumber\\
&-\frac{\mu}{2}\left\|\Theta_{N+1}-\Theta\right\|^2 - \lambda_N \sum_{s=1}^{d} \sum_{t=1}^{n} \frac{1}{\big|\widehat{\Theta}_{N+1}(s,t)\big|}\left|\Theta(s,t)\right|\nonumber\\
&=\frac{\mu}{2}\big\|(\Theta_{N+1}-\Theta)+\widetilde{\Xi}_{N+1}\big\|^2-\frac{\mu}{2}\left\|\Theta_{N+1}-\Theta\right\|^2\nonumber\\
&+\lambda_N \sum_{s=1}^{d} \sum_{t=1}^{n} \frac{\big|\Theta(s,t)-\widetilde{\Xi}_{N+1}(s,t)\big|-\left|\Theta(s,t)\right|}{\big|\widehat{\Theta}_{N+1}(s,t)\big|}\nonumber\\
&\triangleq \triangle F_{N+1}^{(1)}+\triangle F_{N+1}^{(2)}
\end{align}
where
$
\triangle F_{N+1}^{(1)}\triangleq \frac{\mu}{2}\big\|(\Theta_{N+1}-\Theta)+\widetilde{\Xi}_{N+1}\big\|^2-\frac{\mu}{2}\|\Theta_{N+1}-\Theta\|^2
$,
$
\triangle F_{N+1}^{(2)}\triangleq \lambda_N \sum_{s=1}^{d} \sum_{t=1}^{n} \frac{\big|\Theta(s,t)-\widetilde{\Xi}_{N+1}(s,t)\big|-|\Theta(s,t)|}{|\widehat{\Theta}_{N+1}(s,t)|}.
$

For the square term $\triangle F_{N+1}^{(1)} $, by using the inequality $|{\rm tr}(A^\top B)|\leq \|A\| \|B\|$, we have that for some $c_1>0$,
\begin{align}\label{eq:F1}
&\triangle F_{N+1}^{(1)}
= \frac{\mu}{2} {\rm tr}\left(\widetilde{\Xi}_{N+1}^\top \widetilde{\Xi}_{N+1}+2\widetilde{\Xi}_{N+1}^\top (\Theta_{N+1}-\Theta)\right)\nonumber\\
&\geq \frac{\mu}{2} \big\|\widetilde{\Xi}_{N+1}\big\|^2-\mu\big\|\widetilde{\Xi}_{N+1}\big\| \left\|\Theta_{N+1}-\Theta\right\|\nonumber\\
&\geq \frac{\mu}{2} \big\|\widetilde{\Xi}_{N+1}\big\|^2-c_1\mu\big\|\widetilde{\Xi}_{N+1}\big\| \sqrt{\frac{\log\lambda_{\max}(N)}{\lambda_{\min}(N)}}
\end{align}
where for the last inequality Theorem \ref{th:thetan} is applied, i.e., $\left\|\Theta_{N+1}-\Theta\right\|=O\left(\sqrt{\frac{\log\lambda_{\max}(N)}{\lambda_{\min}(N)}}\right)$.

For $\triangle F_{N+1}^{(2)} $, noting that $\mathcal{A}^*$ is the index set of the zero entries of $\Theta$, we have
\begin{align}\label{eq:df(2)}
&\triangle F_{N+1}^{(2)}
= \lambda_N \sum_{(s,t)\in{\mathcal{A}^*}^c} \frac{\big|\Theta(s,t)-\widetilde{\Xi}_{N+1}(s,t)\big|-\left|\Theta(s,t)\right|}{\big|\widehat{\Theta}_{N+1}(s,t)\big|}\nonumber\\
&~~~~~~~~~~~~~~~~~+\lambda_N \sum_{(s,t)\in\mathcal{A}^*} \frac{\big|\widetilde{\Xi}_{N+1}(s,t)\big|}{\big|\widehat{\Theta}_{N+1}(s,t)\big|}\nonumber\\
&\geq \lambda_N \sum_{(s,t)\in{\mathcal{A}^*}^c} \frac{\big|\Theta(s,t)-\widetilde{\Xi}_{N+1}(s,t)\big|-\left|\Theta(s,t)\right|}{\big|\widehat{\Theta}_{N+1}(s,t)\big|}.
\end{align}

Since $\big|\widehat{\Theta}_{N+1}(s,t)\big|\geq |\Theta(s,t)|- \big|\widehat{\Theta}_{N+1}(s,t)-\Theta(s,t)\big|$, $\big|\widehat{\Theta}_{N+1}(s,t)-\Theta(s,t)\big|\mathop{\longrightarrow}\limits_{N\to\infty}0$ and $\left|\Theta(s,t)\right|>0$ for any $(s,t)\in{\mathcal{A}^*}^c$, there exists some integer $N_0>0$ sufficiently large and constants $c_2>0$ and $c_3>0$ such that for any $N\geq N_0$,
\begin{align}
&\left|\lambda_N \sum_{(s,t)\in{\mathcal{A}^*}^c} \frac{\big|\Theta(s,t)-\widetilde{\Xi}_{N+1}(s,t)\big|-\left|\Theta(s,t)\right|}{\big|\widehat{\Theta}_{N+1}(s,t)\big|}\right|\nonumber\\
&\leq c_2 \lambda_N \big\|\widetilde{\Xi}_{N+1}\big\|_1\leq c_3 \lambda_N \big\|\widetilde{\Xi}_{N+1}\big\|
\end{align}
where the last inequality comes from the compatibility of vector norms. Hence it yields that
\begin{align}\label{eq:F2}
\triangle F_{N+1}^{(2)}\geq -c_3 \lambda_N \big\|\widetilde{\Xi}_{N+1}\big\|.
\end{align}

Substituting (\ref{eq:F1}) and (\ref{eq:F2}) into (\ref{eq:dF}), we obtain that
\begin{align}
&0\geq \frac{\mu}{2} \big\|\widetilde{\Xi}_{N+1}\big\|^2-c_1\mu \big\|\widetilde{\Xi}_{N+1}\big\| \sqrt{\frac{\log\lambda_{\max}(N)}{\lambda_{\min}(N)}}\nonumber\\
&~~~~~~-c_3 \lambda_N \big\|\widetilde{\Xi}_{N+1}\big\|\nonumber\\
&= \frac{\mu}{2} \big\|\widetilde{\Xi}_{N+1}\big\|\left(\big\|\widetilde{\Xi}_{N+1}\big\|-2c_1 \sqrt{\frac{\log\lambda_{\max}(N)}{\lambda_{\min}(N)}}-\frac{2c_3}{\mu} \lambda_N\right).\nonumber
\end{align}

Further, by Assumption \ref{assum:recoe} that $\lambda_N=O \left(\sqrt{\frac{\log\lambda_{\max}(N)}{\lambda_{\min}(N)}}\right)$, we get
\begin{align}\label{eq:step2M}
\big\|\widetilde{\Xi}_{N+1}\big\|&=\left\|\Theta-\Xi_{N+1}\right\|\leq 2c_1 \sqrt{\frac{\log\lambda_{\max}(N)}{\lambda_{\min}(N)}}+\frac{2c_3}{\mu} \lambda_N\nonumber\\
&=O\left(\sqrt{\frac{\log\lambda_{\max}(N)}{\lambda_{\min}(N)}}\right),
\end{align}
and thus (\ref{eq:xierr}) is proved.

Noting Assumption \ref{assum:3}, there exists an $\omega$-set $\Omega_{0}$ with $\mathbb{P}\{\Omega_{0}\}=1$ such that
$
\lambda_{\min }(N) \mathop{\longrightarrow} \limits_{N \rightarrow \infty} \infty$ and $\frac{\log\lambda_{\max}(N)}{\lambda_{\min}(N)} \mathop{\longrightarrow} \limits_{N \rightarrow \infty} 0
$
\textcolor{black}{hold} for any $\omega \in \Omega_{0}$. \textcolor{black}{In the following,} we will analyze the estimation sequence on a fixed sample path $\omega\in\Omega_0$.

Recall that $\Xi_{N+1}=\Theta-\widetilde{\Xi}_{N+1}$. Define
\begin{align}
\widetilde{\overline{\Xi}}_{N+1}=
\left\{
\begin{matrix}
\widetilde{\Xi}_{N+1}(s,t), &\text{~~if } (s,t) \in {\mathcal{A}^*}^c,\\
0, &\text{ if } (s,t) \in \mathcal{A}^*,
\end{matrix}
\right.
\end{align}
and $\overline{\Xi}_{N+1}\triangleq\Theta-\widetilde{\overline{\Xi}}_{N+1}$.

By the result that $\|\Xi_N-\Theta\|\to 0$ as $N\to \infty$, for (\ref{eq26}) it suffices to prove that there exists an integer $N_0(\omega)>0$ large enough such that
$
\widetilde{\Xi}_{N+1}(s,t)=0,~\forall (s,t) \in  \mathcal{A}^*,~\forall N>N_0(\omega).
$

Otherwise, there exists $(s_0,t_0)\in \mathcal{A}^*$ and a subsequence $\big\{\widetilde{\Xi}_{N_m+1}\big\}_{m\geq 1}$ such that $\widetilde{\Xi}_{N_m+1}\left(s_0,t_0\right)\neq 0,~m\geq 1$.

By the optimality of $\Xi_{N+1}$, we have
\begin{align}\label{eq:dfzero}
&0\geq F_{N+1}(\Xi_{N+1})-F_{N+1}\big(\overline{\Xi}_{N+1}\big)\nonumber\\
&=\frac{\mu}{2}\big\|\Theta_{N+1}\!-\!\big(\Theta\!-\!\widetilde{\Xi}_{N+1}\big)\big\|^2\!
-\!\frac{\mu}{2}\big\|\Theta_{N+1}\!-\!\big(\Theta\!-\!\widetilde{\overline{\Xi}}_{N+1}\big)\big\|^2\nonumber\\
&+\lambda_N \!\!\sum_{s=1}^d \! \sum_{t=1}^n \!\! \frac{\big|\Theta(s,t)-\widetilde{\Xi}_{N+1}(s,t)\big|\!-\!\big|\Theta(s,t)-\widetilde{\overline{\Xi}}_{N+1}(s,t)\big|}{\big|\widehat{\Theta}_{N+1}(s,t)\big|}\nonumber\\
\nonumber\\
&\triangleq \triangle F_{N+1}^{(1)}+\triangle F_{N+1}^{(2)}
\end{align}
where
$
\triangle F_{N+1}^{(1)} \triangleq \frac{\mu}{2}\big\|\Theta_{N+1}-\big(\Theta-\widetilde{\Xi}_{N+1}\big)\big\|^2-\frac{\mu}{2}\big\|\Theta_{N+1}-\big(\Theta-\widetilde{\overline{\Xi}}_{N+1}\big)\big\|^2,
$
$
\triangle F_{N+1}^{(2)} \triangleq \lambda_N \sum\limits_{s=1}^d \sum\limits_{t=1}^n  \frac{\big|\Theta(s,t)-\widetilde{\Xi}_{N+1}(s,t)\big|}{\big|\widehat{\Theta}_{N+1}(s,t)\big|}
-\lambda_N \sum\limits_{s=1}^d \sum\limits_{t=1}^n \frac{\big|\Theta(s,t)-\widetilde{\overline{\Xi}}_{N+1}(s,t)\big|}{|\widehat{\Theta}_{N+1}(s,t)|}.
$

For the first term $\triangle F_{N+1}^{(1)}$, we obtain that
\begin{align}
&\triangle F_{N+1}^{(1)}
=\frac{\mu}{2}\big\|\big(\Theta_{N+1}-\Theta+\widetilde{\overline{\Xi}}_{N+1}\big)+\big(\widetilde{\Xi}_{N+1}-\widetilde{\overline{\Xi}}_{N+1}\big)\big\|^2\nonumber\\
&~~~~~~~~~~~~~~-\frac{\mu}{2}\big\|\Theta_{N+1}-\Theta+\widetilde{\overline{\Xi}}_{N+1}\big\|^2\nonumber\\
&=\frac{\mu}{2}{\rm tr}\Big(\big(\widetilde{\Xi}_{N+1}-\widetilde{\overline{\Xi}}_{N+1}\big)^\top \big(\widetilde{\Xi}_{N+1}-\widetilde{\overline{\Xi}}_{N+1}\big)\nonumber\\
&~~~~+2\big(\widetilde{\Xi}_{N+1}-\widetilde{\overline{\Xi}}_{N+1}\big)^\top \big(\Theta_{N+1}-\Theta+\widetilde{\overline{\Xi}}_{N+1}\big)\Big)\nonumber\\
&\geq \frac{\mu}{2}\big\|\widetilde{\Xi}_{N+1}-\widetilde{\overline{\Xi}}_{N+1}\big\|^2\nonumber\\
&~~~~-\mu\big\|\widetilde{\Xi}_{N+1}-\widetilde{\overline{\Xi}}_{N+1}\big\| \big\|\Theta_{N+1}-\Theta+\widetilde{\overline{\Xi}}_{N+1}\big\|\nonumber\\
&\geq \frac{\mu}{2}\big\|\widetilde{\Xi}_{N+1}-\widetilde{\overline{\Xi}}_{N+1}\big\|^2\nonumber\\
&~~~~-\mu\big\|\widetilde{\Xi}_{N+1}-\widetilde{\overline{\Xi}}_{N+1}\big\| \big(\big\|\Theta_{N+1}-\Theta\big\|+\big\|\widetilde{\overline{\Xi}}_{N+1}\big\|\big).
\end{align}

By Theorem \ref{th:thetan},
$
\left\|\Theta_{N+1}-\Theta\right\|
=O\left(\sqrt{\frac{\log\lambda_{\max}(N)}{\lambda_{\min}(N)}}\right).
$
Recall the definitions of $\widetilde{\Xi}_{N+1}$ and $\widetilde{\overline{\Xi}}_{N+1}$, by (\ref{eq:step2M}) we obtain
$
\big\|\widetilde{\overline{\Xi}}_{N+1}\big\|
\leq \big\|\widetilde{\Xi}_{N+1}\big\|
=O\left(\sqrt{\frac{\log\lambda_{\max}(N)}{\lambda_{\min}(N)}}\right).
$

Then there exists a constant $c_4>0$ such that
\begin{align}\label{eq:Fbar1}
\triangle F_{N+1}^{(1)}
&\geq \frac{\mu}{2}\big\|\widetilde{\Xi}_{N+1}-\widetilde{\overline{\Xi}}_{N+1}\big\|^2\nonumber\\
&~~~~-c_4 \big\|\widetilde{\Xi}_{N+1}-\widetilde{\overline{\Xi}}_{N+1}\big\| \sqrt{\frac{\log\lambda_{\max}(N)}{\lambda_{\min}(N)}}.
\end{align}

For the second term $\triangle F_{N+1}^{(2)}$, from the definitions of $\widetilde{\Xi}_{N+1}$ and $\widetilde{\overline{\Xi}}_{N+1}$ and noting $\widetilde{\Xi}_{N+1}(s,t)=\widetilde{\overline{\Xi}}_{N+1}(s,t)$ for any $(s,t)\in {\mathcal{A}^*}^c$, we get that
\begin{align}\label{eq:bardf2}
&\triangle F_{N+1}^{(2)}
= \lambda_N \sum_{(s,t)\in \mathcal{A}^*} \frac{\big|\widetilde{\Xi}_{N+1}(s,t)\big|-\big|\widetilde{\overline{\Xi}}_{N+1}(s,t)\big|}{\big|\widehat{\Theta}_{N+1}(s,t)\big|}\nonumber\\
&+\lambda_N \!\!\!\!\!\!\sum_{(s,t)\in {\mathcal{A}^*}^c}\!\!\!\!\!\! \frac{\big|\Theta(s,t)-\widetilde{\Xi}_{N+1}(s,t)\big|-\big|\Theta(s,t)-\widetilde{\overline{\Xi}}_{N+1}(s,t)\big|}{\big|\widehat{\Theta}_{N+1}(s,t)\big|}\nonumber\\
&= \lambda_N \sum_{(s,t)\in \mathcal{A}^*} \frac{\big|\widetilde{\Xi}_{N+1}(s,t)\big|}{\big|\widehat{\Theta}_{N+1}(s,t)\big|}.
\end{align}

From the definition of $\widehat{\Theta}_{N+1}(s,t)$, it yields that for $\forall (s,t)\in \mathcal{A}^*$, there exist constants $c_5>0$ and $c_6>0$ such that
$
c_5 \sqrt{\frac{\log\lambda_{\max}(N)}{\lambda_{\min}(N)}}\leq \big|\widehat{\Theta}_{N+1}(s,t)\big|\leq c_6 \sqrt{\frac{\log\lambda_{\max}(N)}{\lambda_{\min}(N)}}
$
which implies
\begin{align}\label{eq:thetahatabsbound}
\frac{1}{c_6\sqrt{\frac{\log\lambda_{\max}(N)}{\lambda_{\min}(N)}}}
\!\!\leq \!\!\frac{1}{|\widehat{\Theta}_{N+1}(s,t)|}
\!\!\leq \!\!\frac{1}{c_5\sqrt{\frac{\log\lambda_{\max}(N)}{\lambda_{\min}(N)}}}.
\end{align}

Hence for $\triangle F_{N+1}^{(2)}$, \textcolor{black}{from (\ref{eq:bardf2}) and (\ref{eq:thetahatabsbound})} and noting $\widetilde{\Xi}_{N+1}(s,t)=\widetilde{\overline{\Xi}}_{N+1}(s,t)$ \textcolor{black}{for any $(s,t)\in {\mathcal{A}^*}^c$,} we have
\begin{align}\label{eq:Fbar2}
\triangle F_{N+1}^{(2)}
&\geq \frac{1}{c_6} \frac{\lambda_N}{\sqrt{\frac{\log\lambda_{\max}(N)}{\lambda_{\min}(N)}}}  \sum_{(s,t) \in \mathcal{A}^*} \left|\widetilde{\Xi}_{N+1}(s,t)\right|\nonumber\\
&= \frac{1}{c_6} \frac{\lambda_N}{\sqrt{\frac{\log\lambda_{\max}(N)}{\lambda_{\min}(N)}}}  \left\|\widetilde{\Xi}_{N+1}-\widetilde{\overline{\Xi}}_{N+1}\right\|_1 \nonumber\\
&\geq c \frac{\lambda_N}{\sqrt{\frac{\log\lambda_{\max}(N)}{\lambda_{\min}(N)}}}  \left\|\widetilde{\Xi}_{N+1}-\widetilde{\overline{\Xi}}_{N+1}\right\|.
\end{align}

Combining equations (\ref{eq:dfzero}), (\ref{eq:Fbar1}), and (\ref{eq:Fbar2}), we obtain that
\begin{align}
0&\geq F_{N+1}\left(\Xi_{N+1}\right)-F_{N+1}\big(\overline{\Xi}_{N+1}\big) \nonumber\\
&\geq \frac{\mu}{2}\big\|\widetilde{\Xi}_{N+1}-\widetilde{\overline{\Xi}}_{N+1}\big\|^2\nonumber\\
&~~~~-c_4 \mu \big\|\widetilde{\Xi}_{N+1}-\widetilde{\overline{\Xi}}_{N+1}\big\| \sqrt{\frac{\log\lambda_{\max}(N)}{\lambda_{\min}(N)}}\nonumber\\
&~~~~+\frac{c \lambda_N}{\sqrt{\frac{\log\lambda_{\max}(N)}{\lambda_{\min}(N)}}} \big\|\widetilde{\Xi}_{N+1}-\widetilde{\overline{\Xi}}_{N+1}\big\|\nonumber\\
&= \frac{\mu}{2}\big\|\widetilde{\Xi}_{N+1}-\widetilde{\overline{\Xi}}_{N+1}\big\|\Bigg(\big\|\widetilde{\Xi}_{N+1}-\widetilde{\overline{\Xi}}_{N+1}\big\|\nonumber\\
&~~~~-2c_4 \sqrt{\frac{\log\lambda_{\max}(N)}{\lambda_{\min}(N)}}+\frac{2c}{\mu} \frac{\lambda_N}{\sqrt{\frac{\log\lambda_{\max}(N)}{\lambda_{\min}(N)}}}\Bigg)\nonumber\\
&= \frac{\mu}{2}\big\|\widetilde{\Xi}_{N+1}-\widetilde{\overline{\Xi}}_{N+1}\big\|\Bigg(\big\|\widetilde{\Xi}_{N+1}-\widetilde{\overline{\Xi}}_{N+1}\big\|\nonumber\\
&~~~~+\Bigg(\frac{2c}{\mu} -2c_4 \frac{\frac{\log\lambda_{\max}(N)}{\lambda_{\min}(N)}}{\lambda_N}\Bigg)\frac{\lambda_N}{\sqrt{\frac{\log\lambda_{\max}(N)}{\lambda_{\min}(N)}}}\Bigg).
\end{align}

By Assumption \ref{assum:recoe} that $\frac{\log\lambda_{\max}(N)}{\lambda_{\min}(N)}=o(\lambda_N)$, we can further obtain 
$
\frac{\frac{\log\lambda_{\max}(N)}{\lambda_{\min}(N)}}{\lambda_N}=o(1).
$

Now we consider the subsequence $\left\{N_m\right\}_{m\geq 1}$ corresponding to $\left(s_0,t_0\right)$. For any given $\varepsilon>0$, there exists an integer $M_0>0$ such that for any $m\geq M_0$,
$
F_{N_m+1}\left(\Xi_{N_m+1}\right)-F_{N_m+1}(\overline{\Xi}_{N_m+1})
\geq \frac{\mu}{2}\big\|\widetilde{\Xi}_{N_m+1}-\widetilde{\overline{\Xi}}_{N_m+1}\big\|\Bigg(\big\|\widetilde{\Xi}_{N_m+1}-\widetilde{\overline{\Xi}}_{N_m+1}\big\|
+\big(\frac{2c}{\mu} -2c_4 \varepsilon\big)\frac{\lambda_{N_m}}{\sqrt{\frac{\log\lambda_{\max}(N_m)}{\lambda_{\min}(N_m)}}}\Bigg).
$

Recall the assumption that $(s_0,t_0)\in \mathcal{A}^*$ and the subsequence $\big\{\widetilde{\Xi}_{N_m+1}\big\}_{m\geq 1}$ with $\widetilde{\Xi}_{N_m+1}\left(s_0,t_0\right)\neq 0,~m\geq 1$. Then we have $\big\|\widetilde{\Xi}_{N_m+1}-\widetilde{\overline{\Xi}}_{N_m+1}\big\|>0$ for all $m\geq 1$. Choosing $\varepsilon>0$ sufficiently small such that $\frac{2c}{\mu} -2c_4 \varepsilon>0$, it yields
$
F_{N_m+1}\left(\Xi_{N_m+1}\right)-F_{N_m+1}(\overline{\Xi}_{N_m+1})> 0
$
which contradicts with the optimality of $\Xi_{N_m+1}$, i.e.,
$
F_{N_m+1}\left(\Xi_{N_m+1}\right)-F_{N_m+1}(\overline{\Xi}_{N_m+1})\leq 0.
$

So far we conclude that $\widetilde{\Xi}_{N+1}(s,t)=0,~\forall (s,t)\in \mathcal{A}^*,~\forall N\geq N_0$ for some $N_0$ large enough. Noting that $\|\Xi_{N+1}-\Theta\|\to0$ as $N\to\infty$, we further have that $\mathcal{A}(\Xi_{N+1})=\mathcal{A}^*,~\forall N\geq N_0$.

This finishes the proof.

\bibliographystyle{plain}        

\end{document}